\documentclass[12pt,a4paper]{article}
\usepackage[cmex10]{amsmath}
\usepackage{float}
\usepackage{graphicx}
\usepackage{array}
\usepackage{color}
\usepackage{fixltx2e}

\begin{document}

\title{Radiation in Bent Asymmetric Coupled Waveguides}

\author{P. Chamorro-Posada\\ 
Departmento de Teor\'{\i}a de la Se\~nal y Comunicaciones e\\
 Ingenier\'{\i}a Telem\'atica, Universidad de Valladolid,\\
 ETSI Telecomunicaci\'on, Paseo Bel\'en 15,\\
 E-47011 Valladolid, Spain.}

\maketitle

\begin{abstract}

A numerical study of the radiation in coupled bent waveguides is presented.  Such arrangements of curved waveguides in reduced radiation-loss configurations can be used to enhance the quality factor of integrated micro-resonators.  3D full vector computations of the complex modal fields (and propagation constants) and the transient propagation effects have been performed.  The results obtained agree qualitatively with former 2D FDTD analyses. At the same time, the calculations presented meet the accuracy requirements for the design of practical implementations that are unattainable with 2D approaches.

\end{abstract}


\section{Introduction}

Racetrack and ring microresonators are essential elements of the existing integrated optics platforms \cite{bogaerts}.  They permit to implement add-drop multiplexers \cite{little1}, optical filters \cite{little2,chamorro11}, optical switches \cite{almeida}, sensors \cite{vos}, modulators \cite{xu}, and they are the building blocks of slow and fast light systems based on coupled resonator waveguides \cite{scheuer,chamorro09}.  

The intrinsic (or unloaded) quality factor $Q$ of an integrated micro-resonator is determined by the propagation loss.  There exist different mechanisms that contribute to the attenuation of the optical field: the intrinsic loss of the waveguide material, the effect of the roughness of the waveguide walls or the curving of waveguide sections.  In high contrast waveguides, the effect of bending can be negligible even for very small radii of curvature.  Such is the case in silicon on insulator (SOI) photonic circuits.  In this platform, on the other hand, the intrinsic propagation losses are relatively high.  The converse situation is found, for instance, in Si\textsubscript{3}N\textsubscript{4}/SiO\textsubscript{2} integrated circuits, where intrinsic losses are very small, but the reduced waveguide core/cladding refractive index contrast relative to that of SOI results in higher radiation losses due to waveguide curvatures.

Radiation losses in micro-resonators can be reduced if a coupled waveguide is added in the bent section, as it has been recently proposed in \cite{chamorro}.  This can be particularly effective in small-to-moderate waveguide index contrast platforms, such as polymer or silicon nitride.  This strategy had been previously exploited in pulley-type microdisk \cite{hosseini,hu} and microring \cite{cai} resonators.  In pulley geometries, the same coupled structure is used for both reducing the radiation and coupling the resonator to an access waveguide, therefore limiting the flexibility in the design when compared with the approach of \cite{chamorro}. The results of this study are relevant both for Q-enhanced \cite{chamorro} and pulley \cite{cai} micro-resonator configurations.

In this work, we present a full vector 3D analysis of the radiation properties of asymmetric coupled waveguides that extend previous 2D studies \cite {chamorro}.  The estimation of the radiation losses in curved structures is often based on 2D approximations that rely on the effective index method (EIM) \cite{kumar,chiang}.  Solutions for the lower dimensional case are then found either using numerical modal \cite{goyal, thyagarajan}, time-domain methods \cite{chamorro}, or with an analytical treatment either exact \cite{hiremath,morita} or approximate \cite{marcuse}, particularly based on the WKB method \cite{berglund, heiblum,kozyreff}. These studies use a scalar or a semivectorial approach, with pure TE or TM polarized fields \cite{chiang}.  Nevertheless, the intrinsic accuracy limitations of the EIM \cite{chiang} are expected to be even deeper when bent geometries are addressed, as it is discussed below. The full 3D vector analyses presented in this work, based on finite-difference approximations, embrace both the properties of lossy modes of the structures and the spatial transients and are free from the limitations of 2D semi-vectorial approaches.

\section{Modeling}\label{sec::modeling}

\begin{figure}[H]
\centering
\includegraphics[width=2.5in]{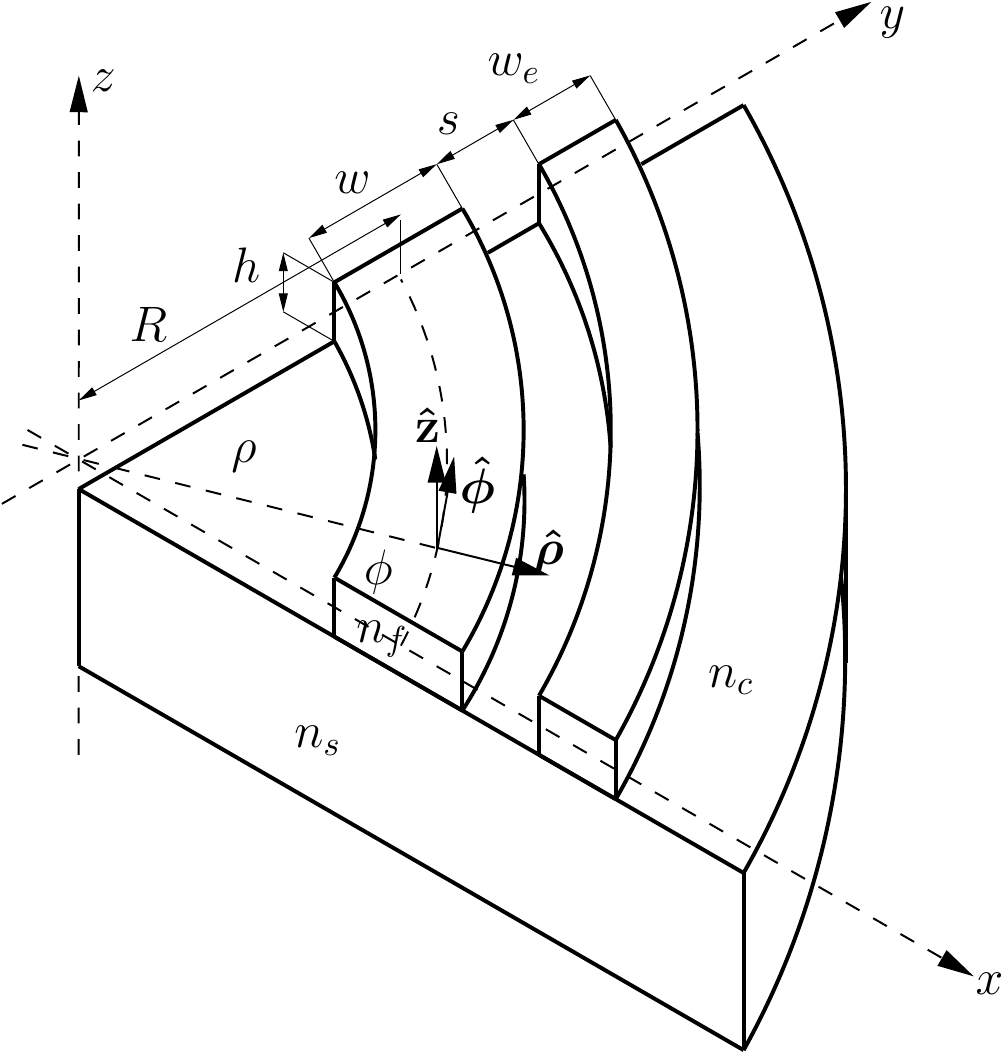}\\
\caption{Geometry of the waveguide structure under study}
\label{fig::geom3D}
\end{figure}

Fig. \ref{fig::geom3D} displays the geometry of the integrated structure under study.  The main transmission system is the inner curved waveguide, with a rectangular section of width $w$ and height $h$. It is illuminated by an input waveguide and it is designed to transmit the optical signal with the lowest possible radiation loss.  To accomplish this goal, it is side-coupled to a second curved waveguide that is also rectangular and has the same height $h$ as the main guide.  Its width $w_e$  and the separation distance $s$ have to be adjusted so as to minimize the radiation.  $n_g$, $n_c$ and $n_s$ are the refractive indices of the guiding, substrate and cover regions.  In all the calculations, we assume an optical carrier wavelength of $\lambda=1550$ nm and the same guiding structure as in \cite{chamorro} with $n_s=n_c=1.4501$ corresponding to SiO\textsubscript{2}  and $n_g=1.9792$ to Si\textsubscript{3}N\textsubscript{4}.  Also, $w=1$ $\mu$m and $h=300$ nm is assumed in all the calculations.  We  focus on the fundamental quasi-TE and quasi-TM modes, with electric fields predominantly polarized along the $\rho$ and $z$ directions, respectively.  Furthermore, we will refer to quasi-TE and quasi-TM modes as TE and TM, respectively, when there is no possibility of confusion.

The propagation medium depicted in Fig. \ref{fig::geom3D} is piecewise homogeneous.  For each of the regions with a given refractive index, both the magnetic field and electric field intensities obey the Helmholtz equation $\nabla\times\nabla\times\left\lbrace\mathbf{E},\mathbf{H}\right\rbrace +k^2n^2\left\lbrace\mathbf{E},\mathbf{H}\right\rbrace=0$.  The electromagnetic field in the structure is also determined by the conditions set by the interfaces.

We seek complex, quasi-guided, mode solutions of the type
\begin{equation}
\left\lbrace\mathbf{E},\mathbf{H}\right\rbrace (\rho,\phi,z)=\left\lbrace\mathbf{e},\mathbf{h}\right\rbrace (\rho,z)\exp\left(j\beta R \phi\right).\label{modo}
\end{equation}
 $\left\lbrace\mathbf{e},\mathbf{h}\right\rbrace (\rho,z)$ are the modal fields and $\beta$ is the complex mode propagation constant.  The imaginary part of $\beta$ accounts for radiation losses.  Even though these solutions have some similarities with leaky modes found in straight geometries, they do not diverge at large $\rho$ and can be normalized to the azimuthal modal power flow \cite{hiremath}.   The complex effective index of the modal field $n=n_r+jn_i$ is related to the propagation constant as $\beta=k(n_r+jn_i)$ with $k=\omega/c$.  

When the ansatz \eqref{modo} is substituted in Maxwell Equations, one obtains the eigenvalue equations for the mode field.  It is only necessary to calculate two field components, which can be used to derive the full electric and magnetic field intensity vectors.  For instance \cite{krause},
\begin{eqnarray}
\left(1+\dfrac{r}{R}\right)^2Lh_r+\dfrac{1}{R}\left(1+\dfrac{r}{R}\right)\left(3\dfrac{\partial h_r}{\partial r}+2\dfrac{\partial h_z}{\partial z}\right)+\\\nonumber
\dfrac{1}{R^2}h_r=\beta^2h_r\\
\left(1+r/R\right)^2Lh_z+ \dfrac{1}{R}\left(1+\dfrac{r}{R}\right)\dfrac{\partial h_z}{\partial r}=\beta^2h_z,
\end{eqnarray} 
with
\begin{equation}
L=\dfrac{\partial^2}{\partial r^2}+\dfrac{\partial^2}{\partial z^2}+k^2 n^2
\end{equation}
and $\rho=R+r$.  This transformation permits to set the $r=0$ reference at the central point of the main waveguide  in the geometry described in Fig. \ref{fig::geom3D}.  The full 3D vector modes can be obtained, for instance, using a finite difference discretization of the above equations  \cite{krause,kim,lui}.  All the 3D vector mode field calculations in this work have been performed using the 
\emph{wgms3d} \cite{krause,wgms3d} software package.

A very common approach to this problem is also based on its reduction to a 2D geometry, using the EIM \cite{kumar,chiang}.  Solutions for the 2D effective problem are then sought for by different means \cite{goyal, thyagarajan,chamorro,hiremath,morita,marcuse,berglund, heiblum,kozyreff}.  In the EIM, a semivectorial approach with pure TE or TM polarized fields is being used \cite{chiang}.  By assuming the separability of both the field components and the refractive index \cite{kumar,chiang}, the problem is first analyzed in one of the transverse spatial dimensions, producing the effective indices for the analysis along the orthogonal transverse direction.

\begin{figure}[H]
\centering
\begin{tabular}{c}
{\large (a)}\\
\includegraphics[width=2.8in]{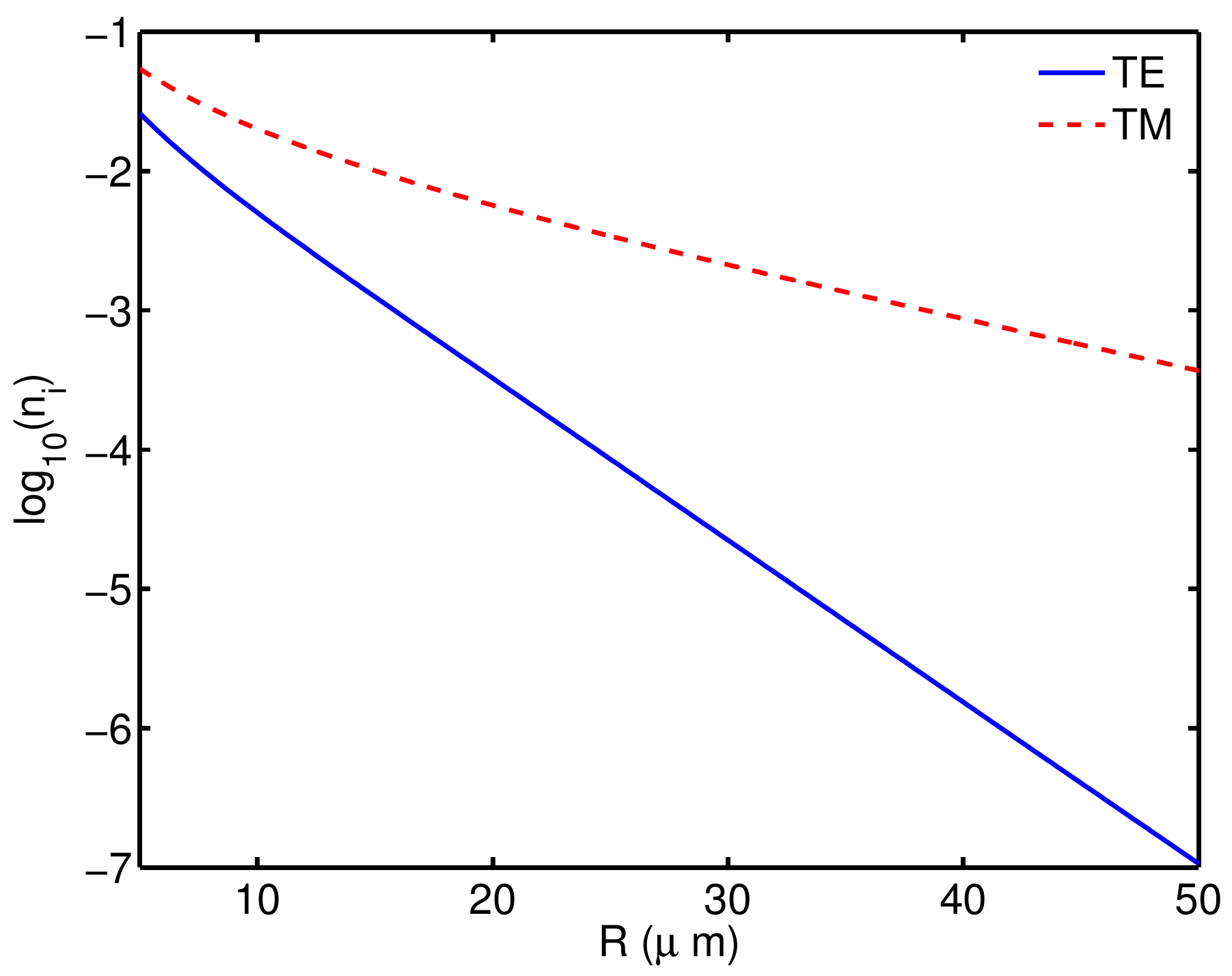}\\
\begin{tabular}{cc}
\includegraphics[width=1.5in]{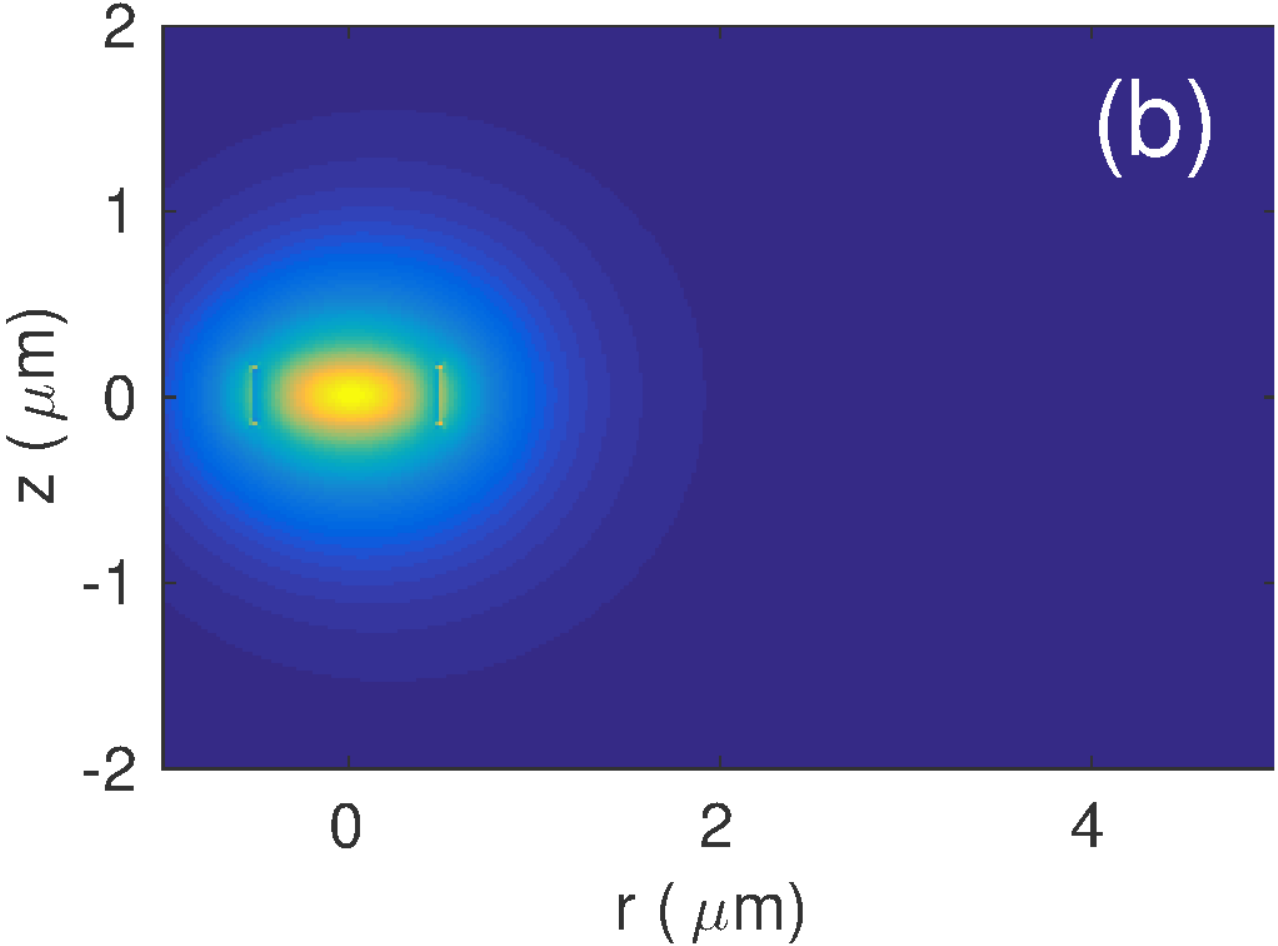}&\includegraphics[width=1.5in]{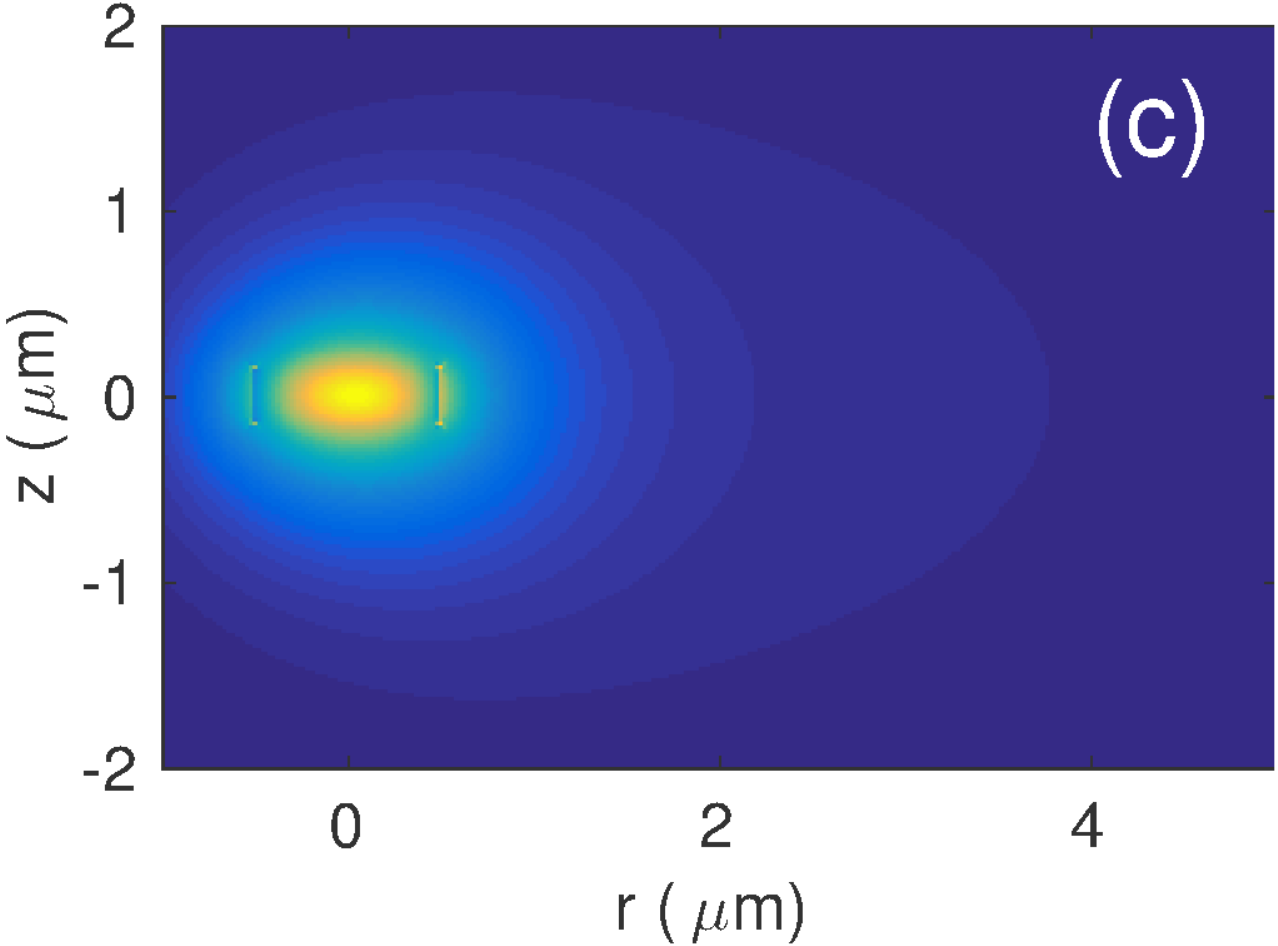}\\
\includegraphics[width=1.5in]{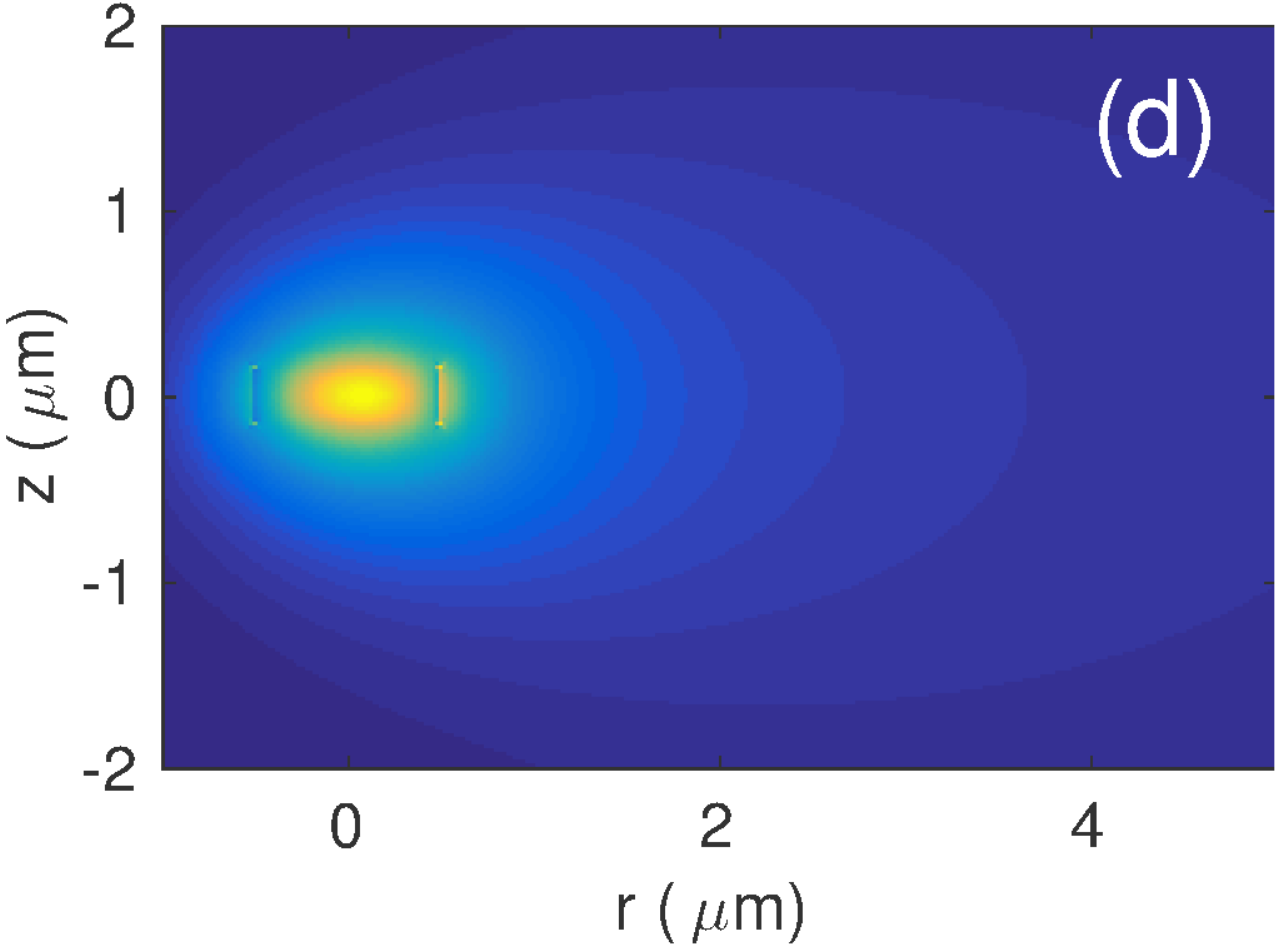}&\includegraphics[width=1.5in]{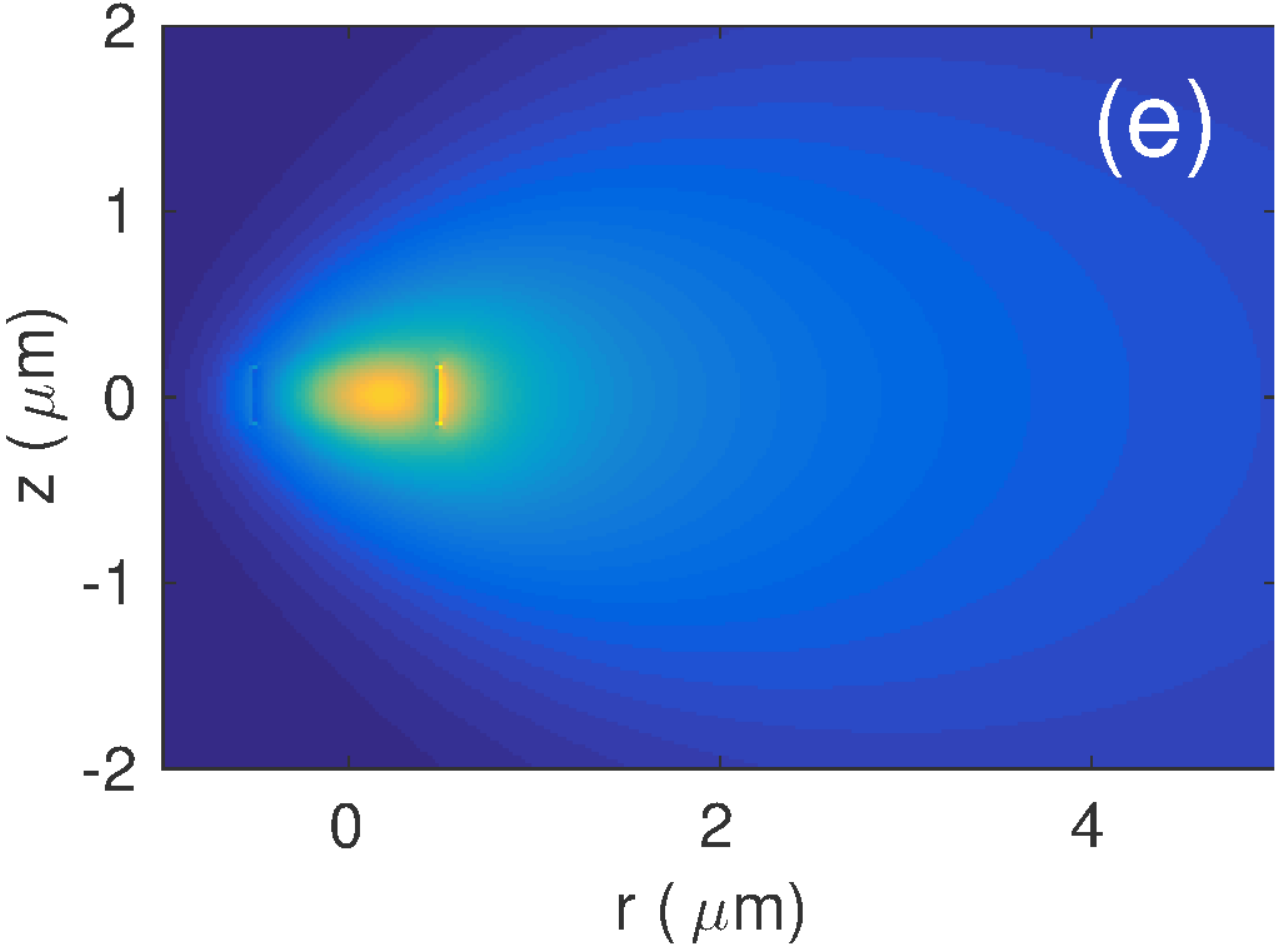}
\end{tabular}
\end{tabular}
\caption{(a) Logarithmic plot of the imaginary effective index of the quasi-TE and quasi-TM modes of the isolated single waveguide. (b) to (e) display transverse field intensity of the modal TE field, predominantly polarized in the $\rho$ direction for  $R$ equal to $50$ $\mu$m, $25$ $\mu$m, $15$ $\mu$m and $5$ $\mu$m, respectively.}
\label{fig::radiacion}
\end{figure}

The intrinsic error in the effective index model \cite{chiang} is due to: (a) the semi-vectorial approach to a full vector problem and (b) the fact that the refractive index is actually non-separable.  Both limitations become more severe for the refractive index differences associated to the silicon and silicon nitride photonic platforms.  In the case of curved waveguides, the modal fields themselves become evidently non-separable, which results in an expected increase of the deviation of the results from the exact solutions.  This intrinsic 3D character of quasi-modes in bent waveguides is illustrated in Figs. \ref{fig::radiacion} $(b)$ to  \ref{fig::radiacion} $(e)$ that  show the transverse electric field intensities for the TE modal fields at four values of the bend radius.  For $R=5$ $\mu$m, it is obvious that the modal field is not separable.  The lack of this condition can also be clearly observed for $R=15$ $\mu$m and $R=25$ $\mu$m, and it is noticeable for $R=50$ $\mu$m.  Fig. \ref{fig::radiacion} (a) displays the logarithm of the imaginary part of the modal index as a function of the bend radius $R$.

The study of the spatial transients in the structure is performed using a full vector beam propagation method.  We define the propagation coordinate as $u=R\phi$, and the electromagnetic field envelope $\left\lbrace\widetilde{\mathbf{E}},\widetilde{\mathbf{H}}\right\rbrace$ as  $\left\lbrace\mathbf{E},\mathbf{H}\right\rbrace=\left\lbrace\widetilde{\mathbf{E}},\widetilde{\mathbf{H}}\right\rbrace\exp\left(jkn_0u\right)$, where $n_0$ is an arbitrary reference refractive index. 

Under the conditions of the slowly varying envelope approximation (SVEA), $\left|2kn_0\dfrac{\partial\widetilde{E}_i}{\partial u}\right|>>\left|\dfrac{\partial^2\widetilde{E}_i}{\partial u^2}\right|$, we can neglect the second $u$ derivative of the optical field. If, also, the refractive index does not vary with $u$, the transverse electric field evolution can be approximated with the paraxial evolution \cite{lui}
\begin{equation}
2jkn_0\dfrac{\partial}{\partial u}
\left[
\begin{matrix}
\widetilde{E}_r\\
\widetilde{E}_z
\end{matrix}
\right]=
\left[
\begin{matrix}
A_{rr} & A_{r z}\\
A_{zr} & A_{zz}\\
\end{matrix}
\right]
\left[
\begin{matrix}
\widetilde{E}_r\\
\widetilde{E}_z
\end{matrix}
\right].\label{eq::bpm}
\end{equation}

If the calculations are restricted to a region close to the waveguide axis $2|r|<<R$, and it can be assumed that the effective wavelength is much shorter than the bending radius $|kn_0R|>>1$ and $\left|\dfrac{\partial}{\partial r}\right|>>\dfrac{3}{R}$, the operators can be approximated as \cite{lui}
\begin{eqnarray}
A_{\alpha\alpha}\widetilde{E}_\alpha=\dfrac{\partial}{\partial\alpha}\left[\dfrac{1}{n_t^2}\dfrac{\partial}{\partial\alpha}\left(n_t^2\widetilde{E}_\alpha\right)\right]+\dfrac{\partial^2\widetilde{E}_\alpha}{\partial \beta^2}+k^2\left(n_t^2-n_0^2\right)\widetilde{E}_\alpha \label{eq::bpm1}\\
A_{\alpha\beta}\widetilde{E}_\beta=\dfrac{\partial}{\partial\alpha}\left[\dfrac{1}{n_t^2}\dfrac{\partial}{\partial\beta}\left(n_t^2\widetilde{E}_\beta\right)\right]-\dfrac{\partial^2\widetilde{E}_\beta}{\partial\alpha\partial\beta},\label{eq::bpm2}
\end{eqnarray}
where $(\alpha,\beta)$ equals either $(r,z)$ or $(z,r)$, and $n_t=\left(1+\dfrac{r}{R}\right)n(r,z)$.  Equations \eqref{eq::bpm}, \eqref{eq::bpm1} and \eqref{eq::bpm2} coincide with those defining the approximate vector propagation of an optical beam in the absence of curvature \cite{huang}.

As in the mode field calculations, finite differences are employed to discretize the operators in \eqref{eq::bpm}, \eqref{eq::bpm1} and \eqref{eq::bpm2}. Whereas the former admit schemes with improved accuracy (e.g. \cite{vassallo,ychiang}), their derivation is based on the assumption of a field corresponds to a mode of the structure.  Therefore, we employ general difference discretizations of the operators \cite{huang} for the analysis of the spatial transients using beam propagation.  Perfectly matched layers are implemented using complex coordinate strectching \cite{chew} as absorbing boundary conditions.

\section{The modes of the straight asymmetric coupled waveguides}

\begin{figure}[H]
\centering
\begin{tabular}{cc}
{ (a)}&{(b)}\\
\includegraphics[width=1.6in]{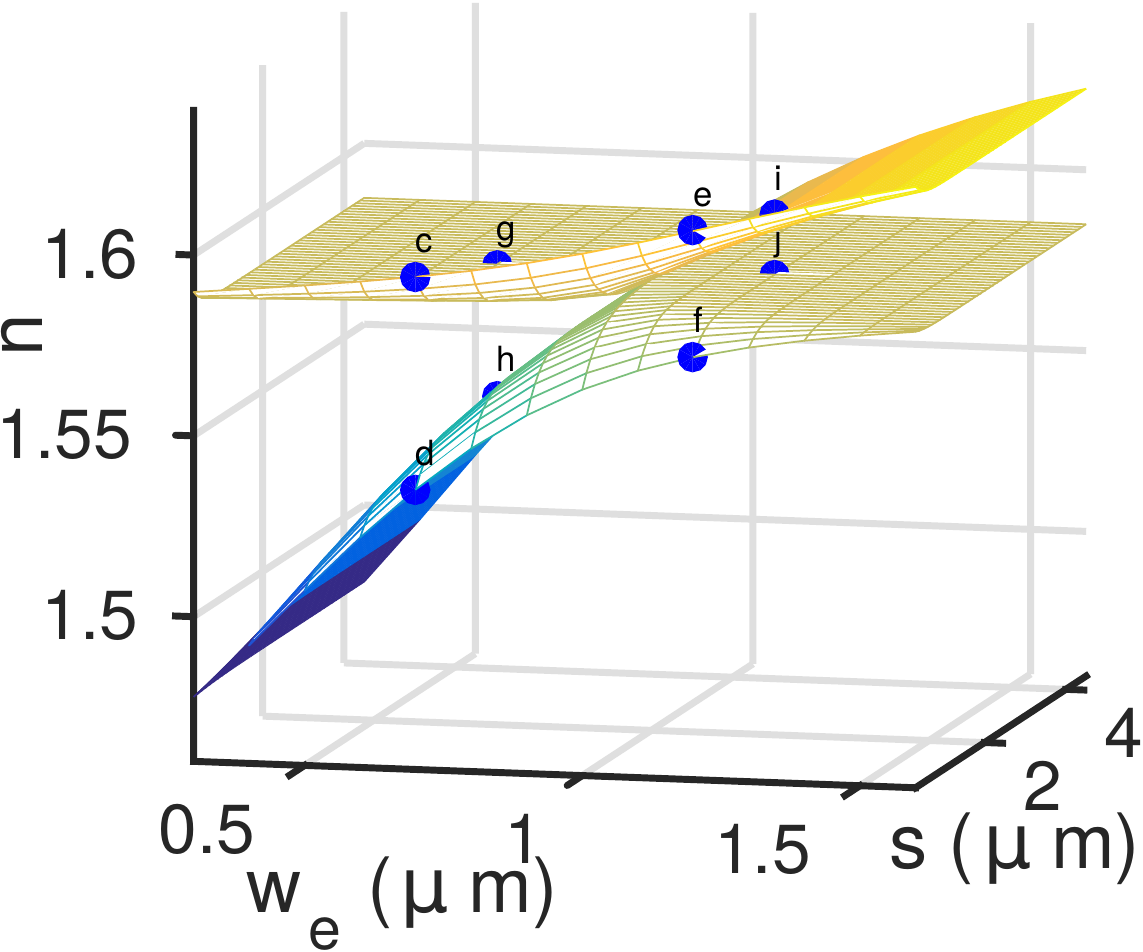}&\includegraphics[width=1.6in]{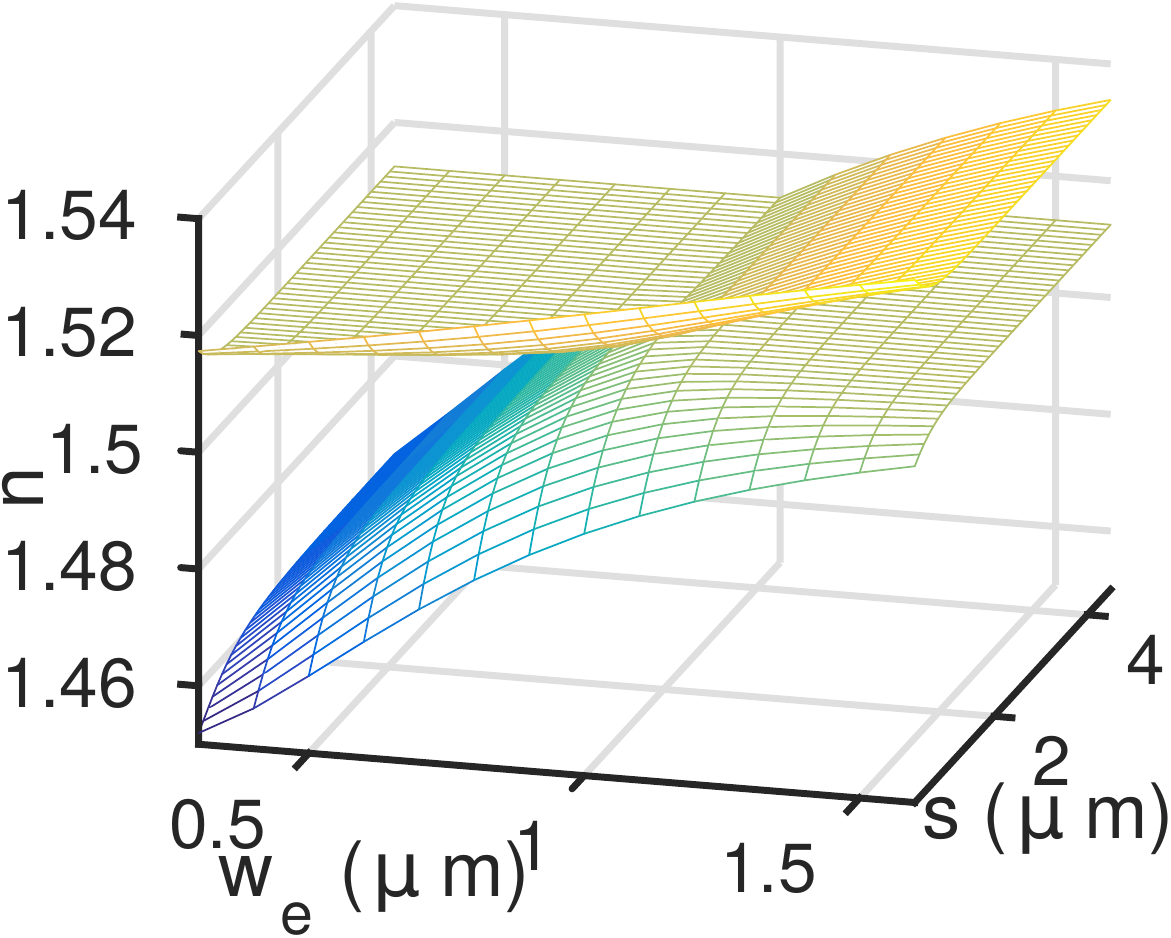}\\
\includegraphics[width=1.6in]{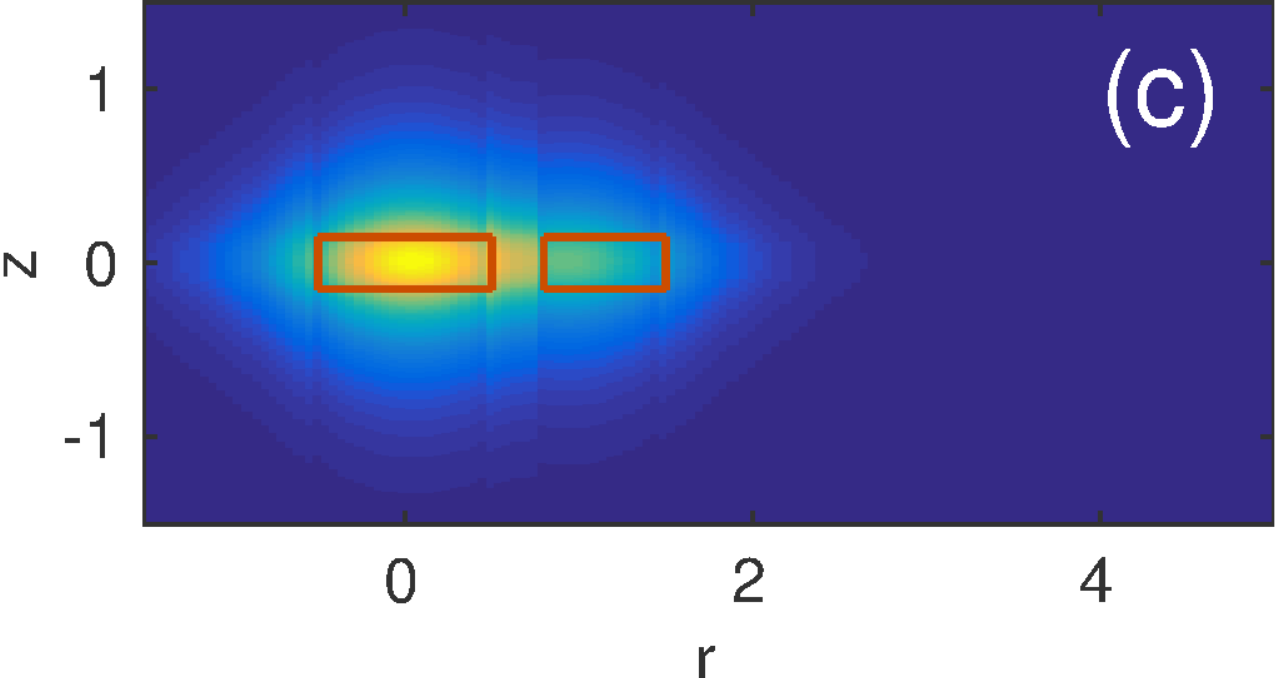}&\includegraphics[width=1.6in]{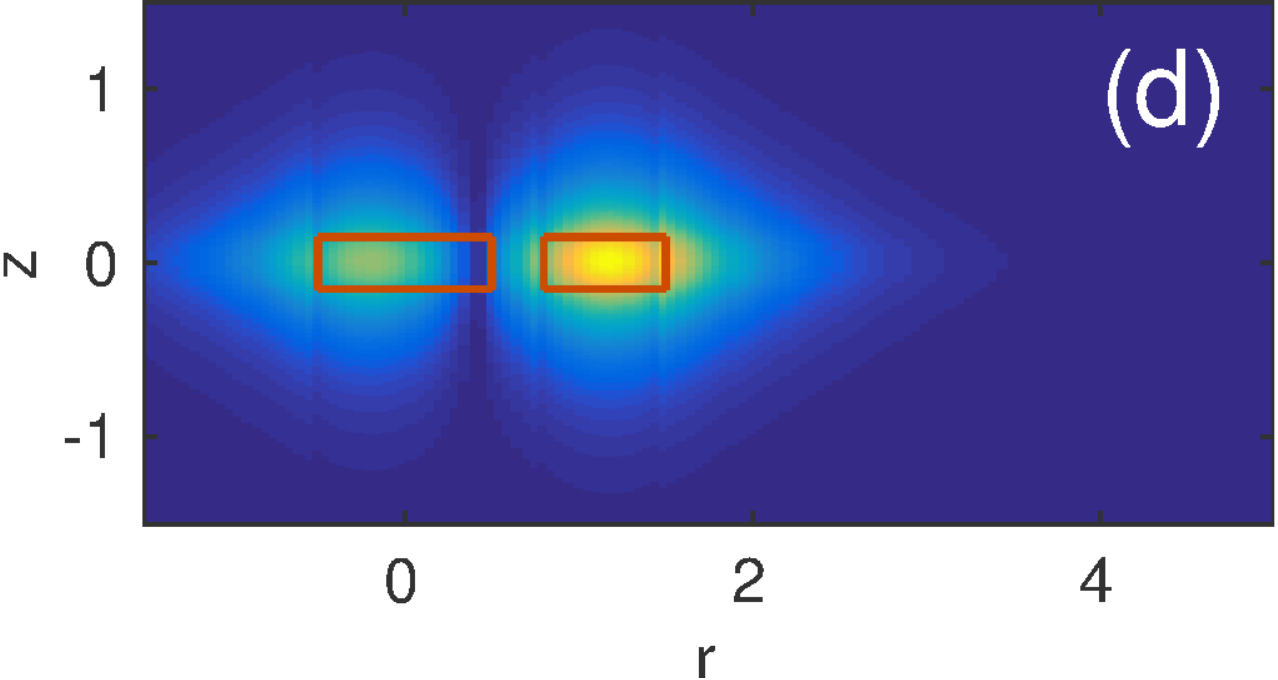}\\
\includegraphics[width=1.6in]{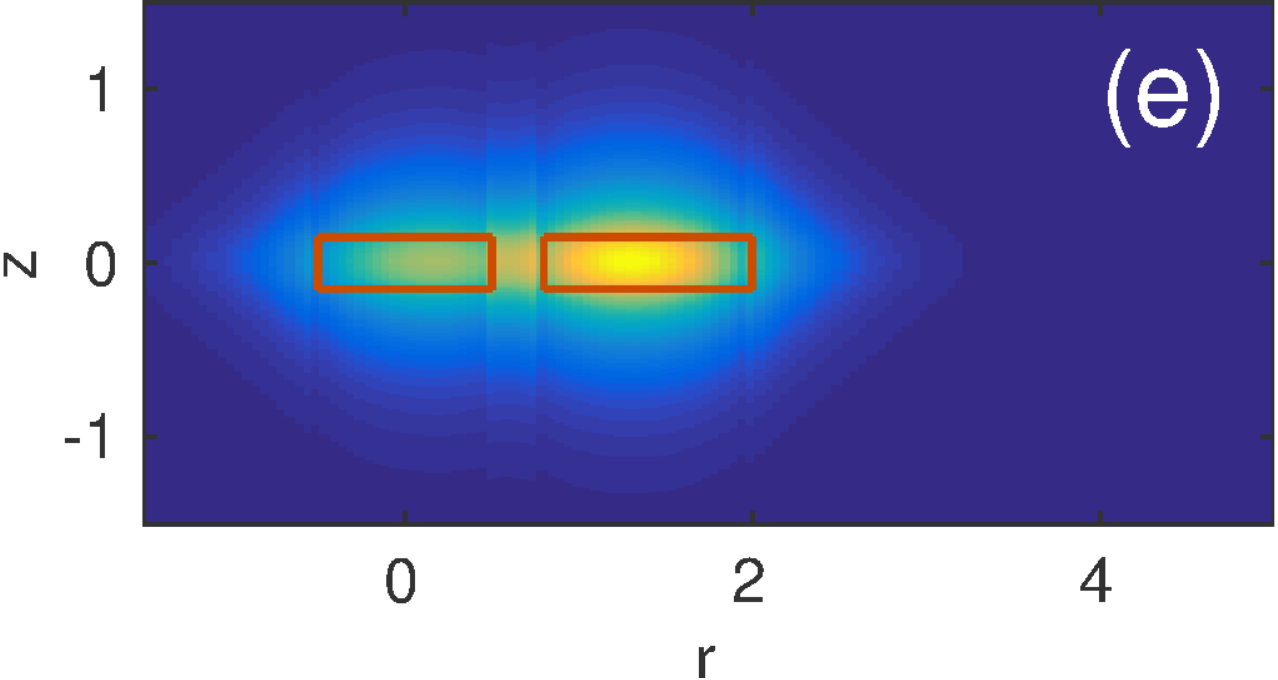}&\includegraphics[width=1.6in]{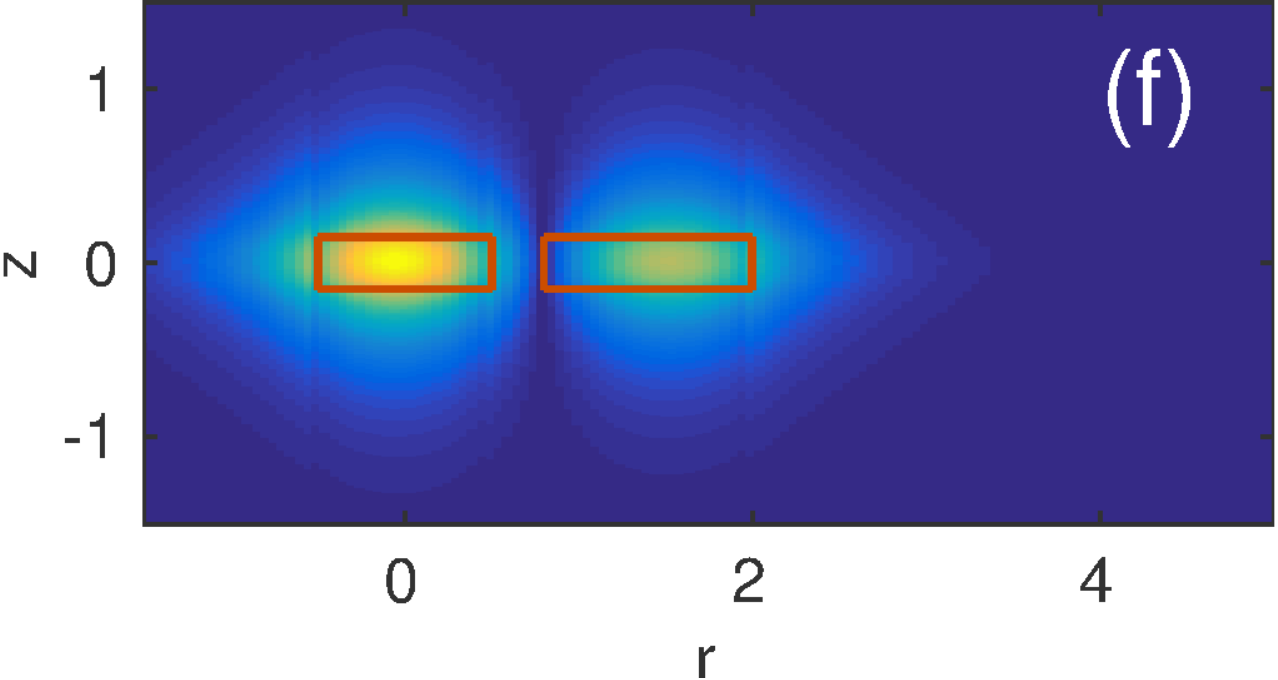}\\
\includegraphics[width=1.6in]{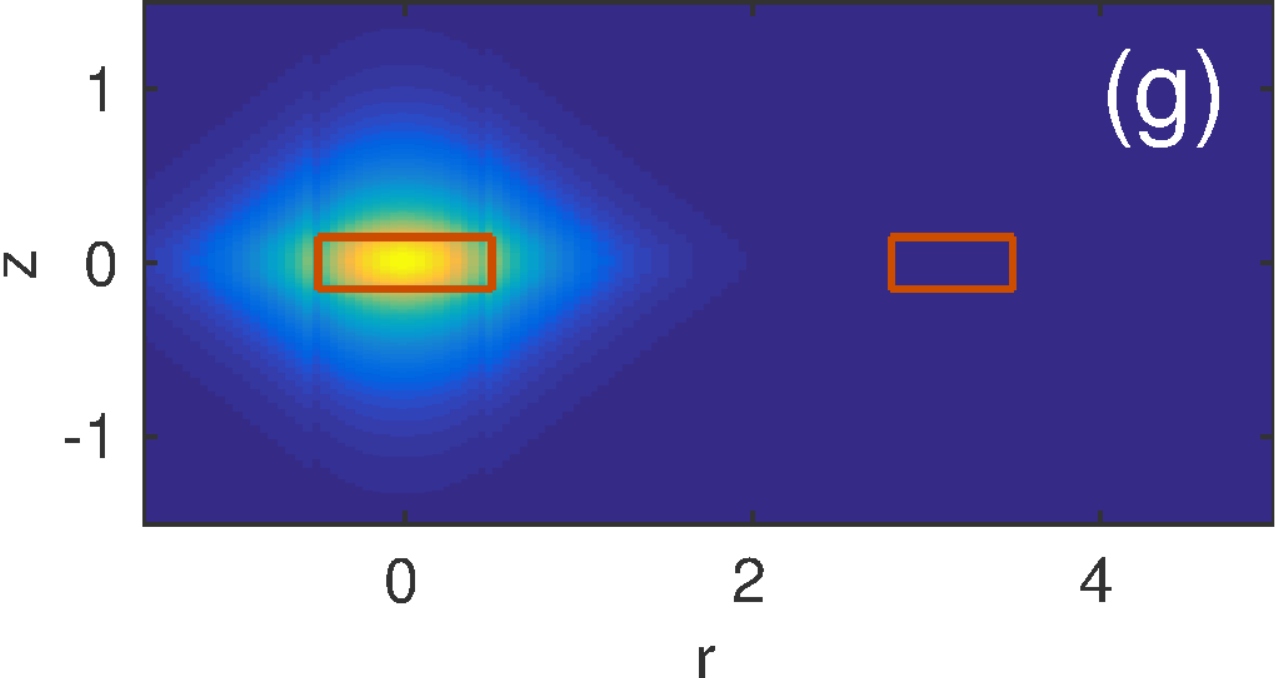}&\includegraphics[width=1.6in]{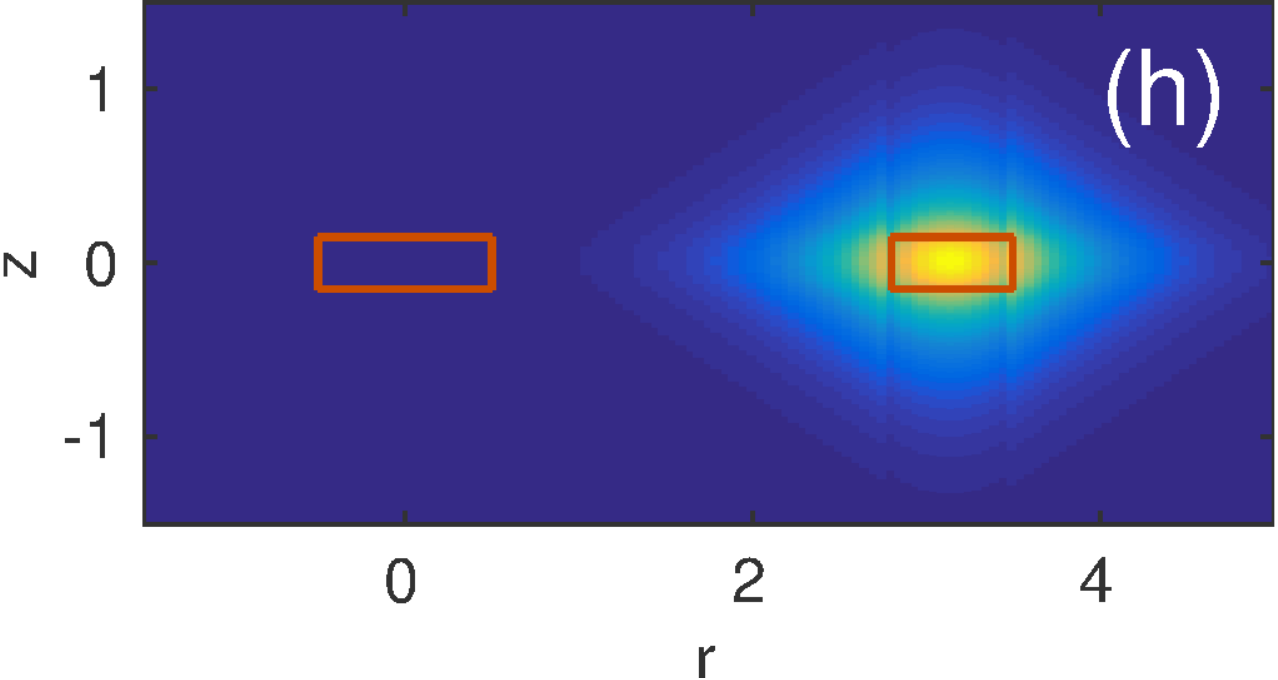}\\
\includegraphics[width=1.6in]{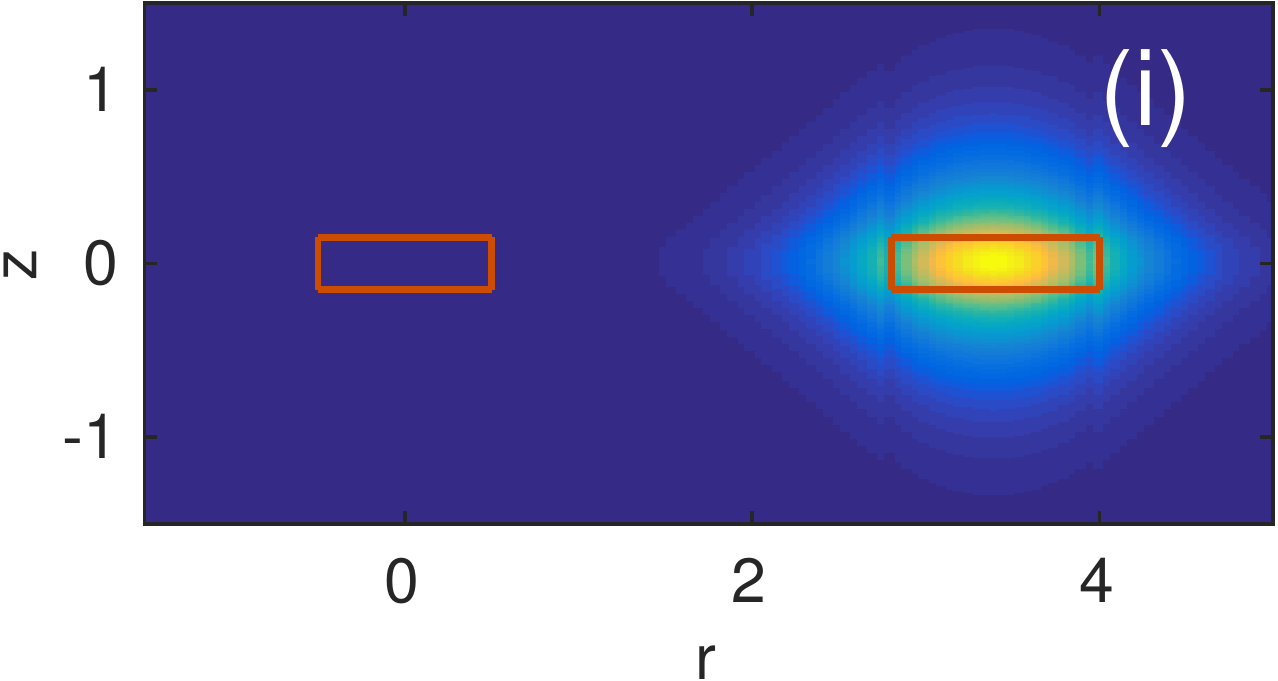}&\includegraphics[width=1.6in]{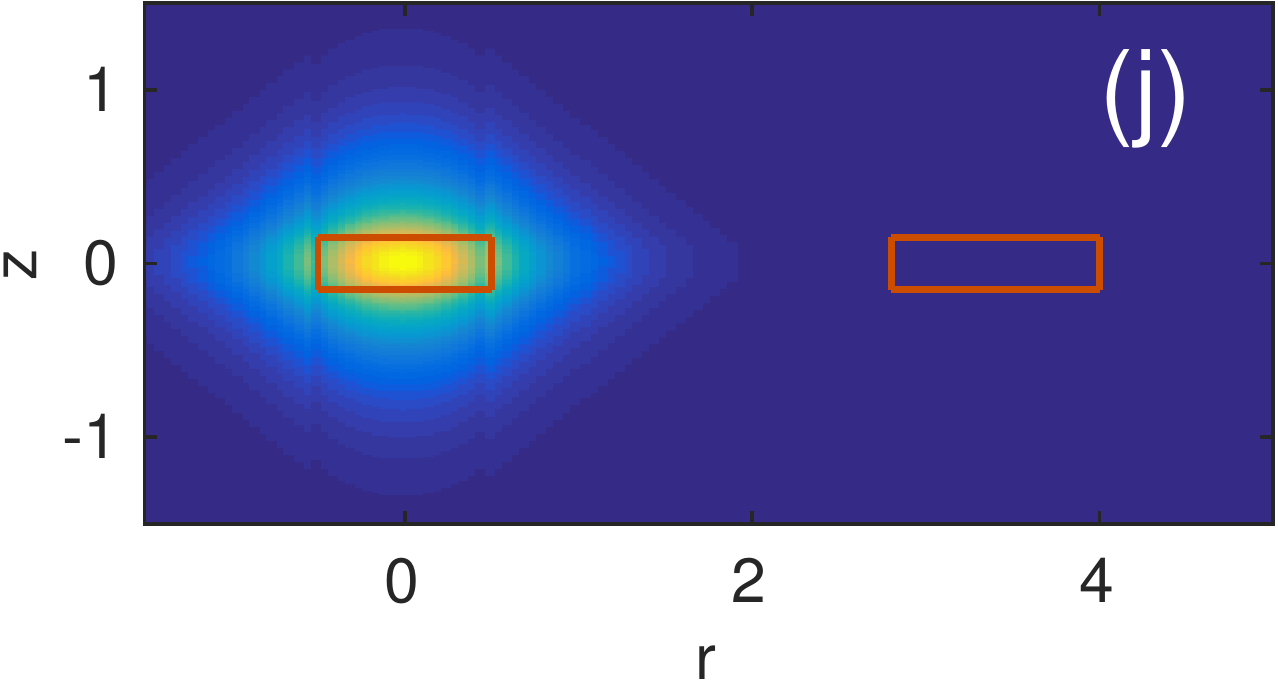}\\
\end{tabular}
\caption{Modal properties of the straight asymmetric coupler calculated with the EIM.  (a) and (b) show the effective indices as a function of the coupler parameters $s$ and $w_e$ for TE and TM polarizations, respectively.  (c) to (f) display the modal fields (transverse electric field intensity) corresponding to the configurations labeled in (a).  Results have been calculated using the EIM.}
\label{fig::rect}
\end{figure}

We consider first the properties of the modes in an  asymmetric coupler in the limit $R\to\infty$.  Since we are only interested in the qualitative description of the dispersion surfaces and the modal fields, we  use the EIM.  Figs. \ref{fig::rect} (a) and \ref{fig::rect} (b) show the approximate modal indices for TE (a) and TM (b) polarized fields.  The parameters used for the device geometry and refractive indices are those previously indicated in Section \ref{sec::modeling}. The results displayed also include those of the symmetric coupler at $w_e=1$ $\mu$m.  In this symmetric case, we observe the mode degeneracy of the uncoupled waveguides at large $s$.  As the guide separation $s$ is reduced, the diagram shows the expected splitting into an even and an odd super-mode that correspond, respectively, to the upper and lower branches of the figure.      For all values of $w_e$ and $s$, the two sheets that describe the modal indices can be obtained as the continuation of the even (upper sheet) and odd (lower sheet) mode values for the symmetric case.

As regards the mode fields, the waveguides are essentially uncoupled for large $s$ and any of the two modes remains localized at either waveguide.  Which mode is actually localized within the main waveguide for large $s$ changes as the $w_e=1$ $\mu$m line, corresponding to a symmetric coupler, is crossed:  It is the upper sheet mode for $w_e<1$ and the lower sheet mode for $w_e>1$.  This is evidenced by the presence of an asymptotic horizontal plane with a modal index close to that of the fundamental mode of the (isolated) main waveguide.  Figs. \ref{fig::rect} (c) to \ref{fig::rect} (d) display the transverse electric field intensities obtained with the EIM for a TE polarized field at specific points in the $(w_e,s)$ plane that have been labeled in Fig. \ref{fig::rect} (a).

\begin{figure}[H]
\centering
\begin{tabular}{cc}
{ (a)}&{(d)}\\
\includegraphics[width=1.7in]{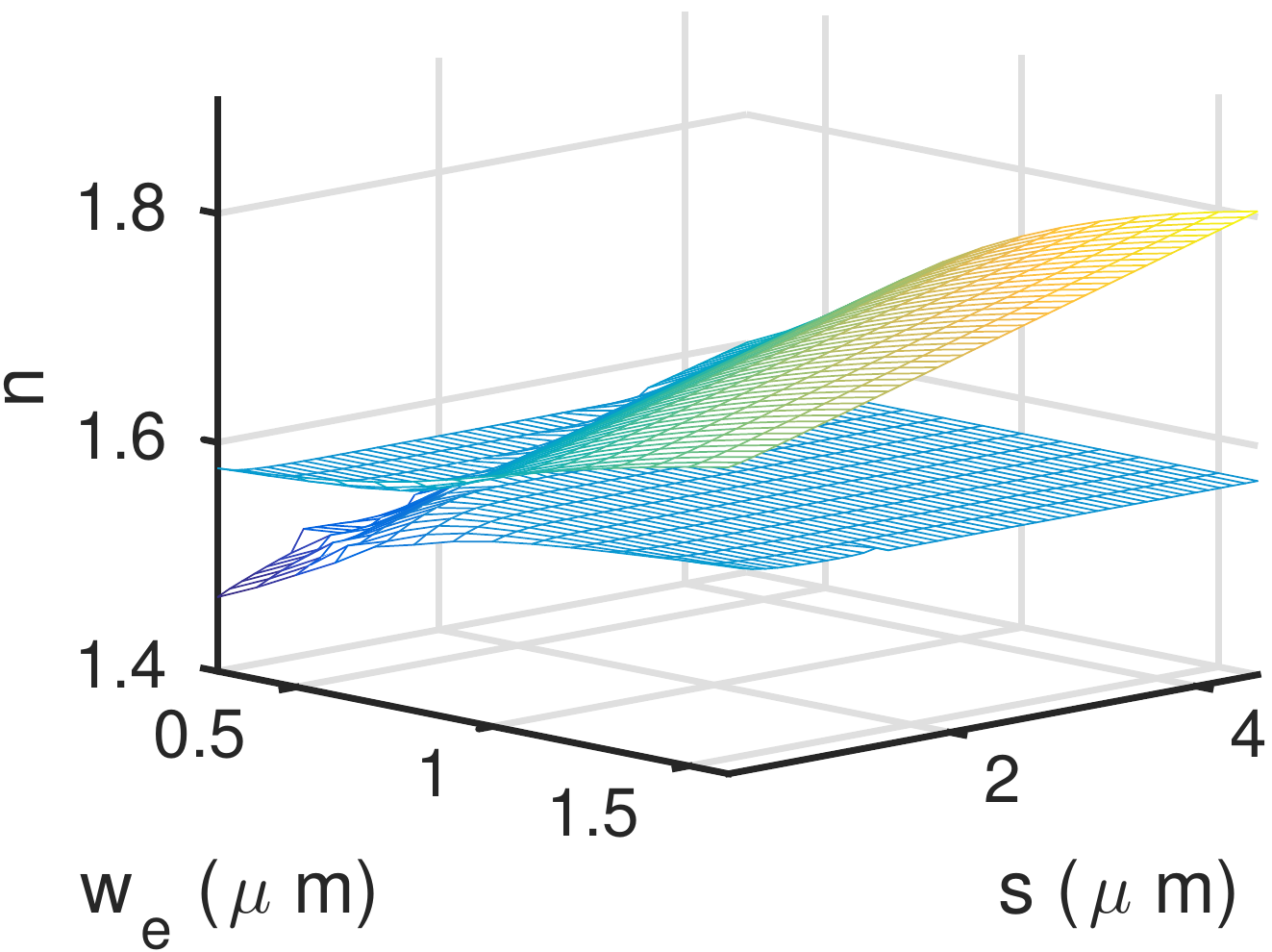}&\hspace*{-0.75cm}
\includegraphics[width=1.7in]{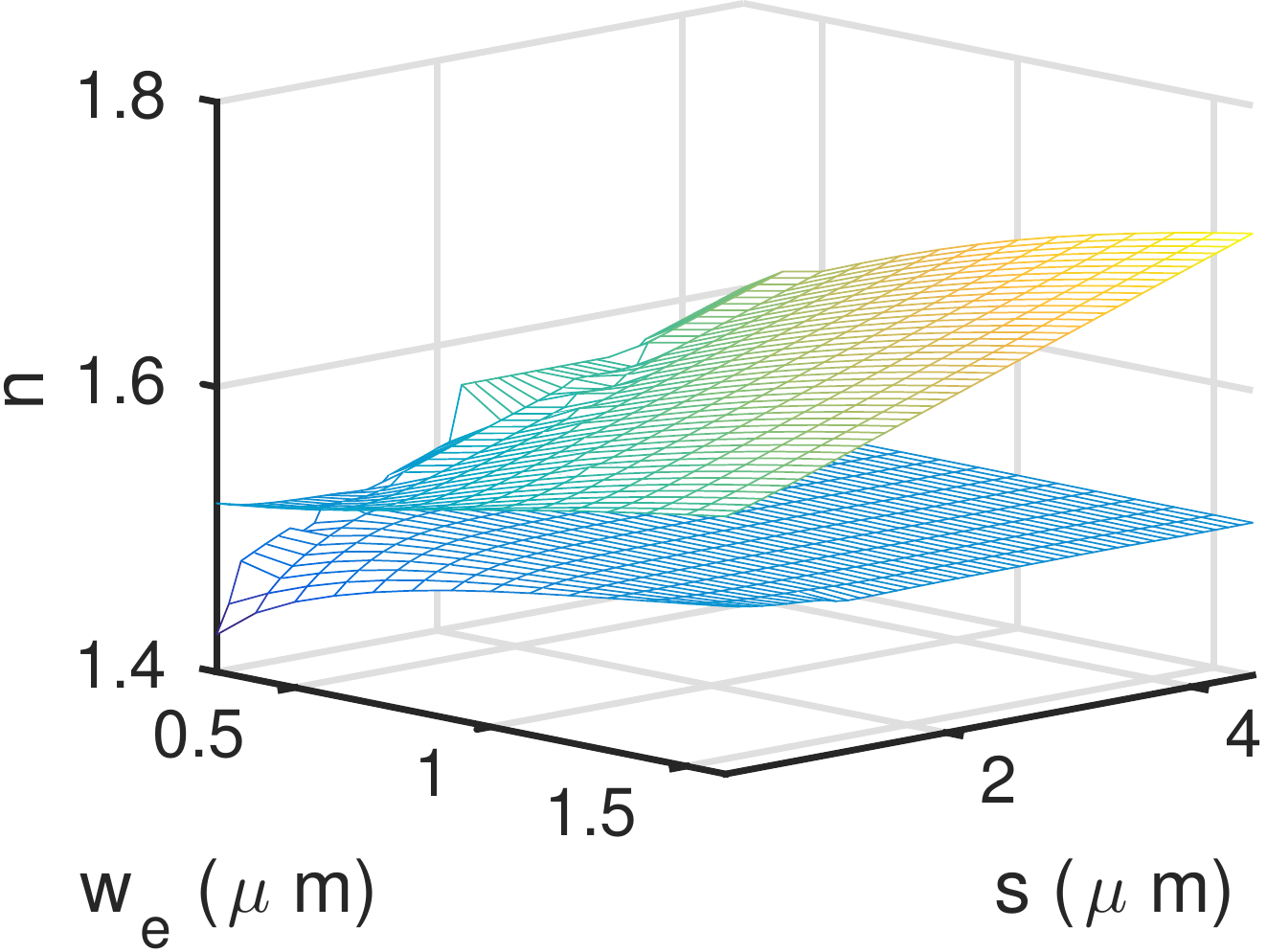}\\
{ (b)}&{(e)}\\
\includegraphics[width=1.7in]{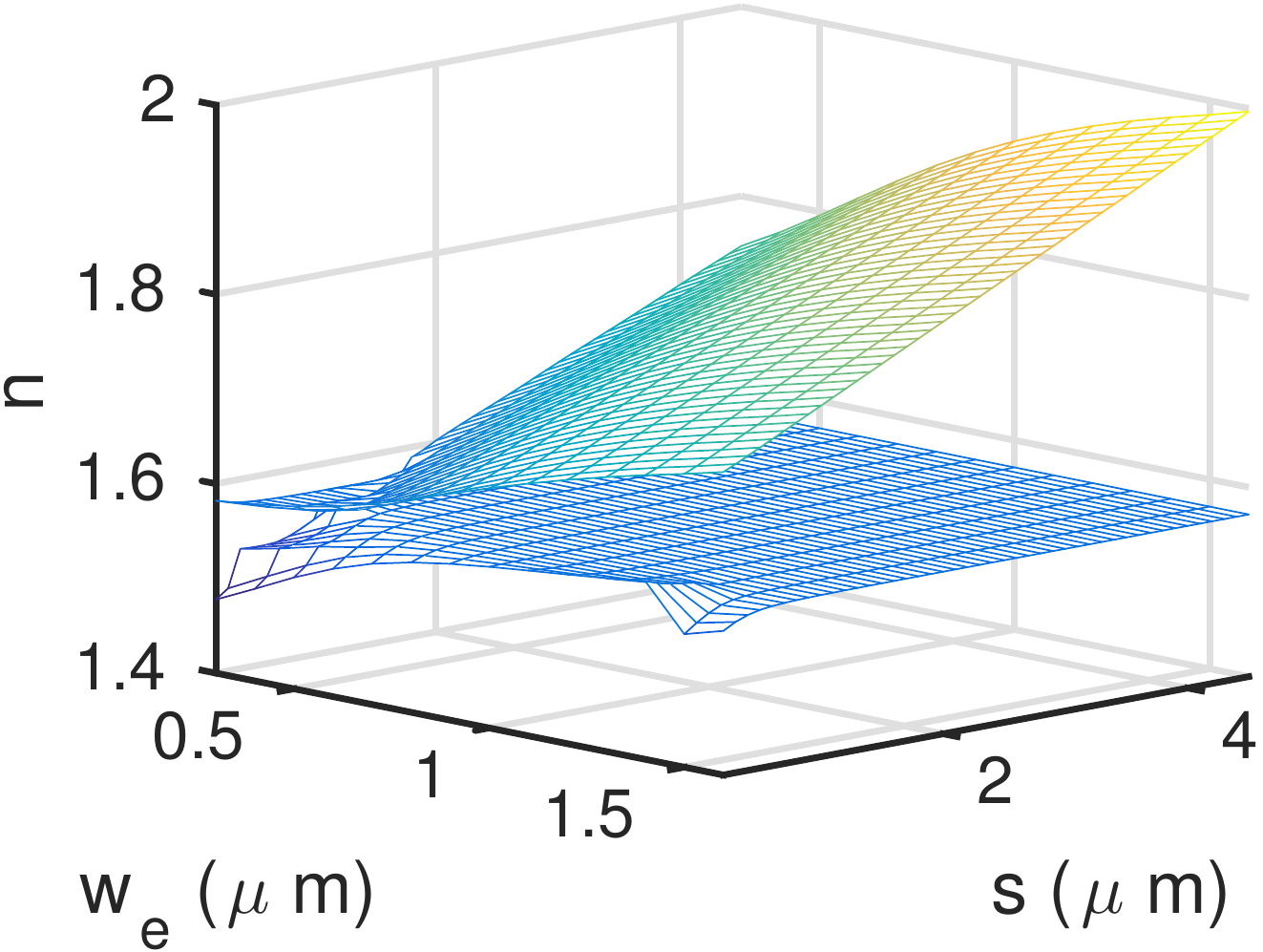}&\hspace*{-0.75cm}
\includegraphics[width=1.7in]{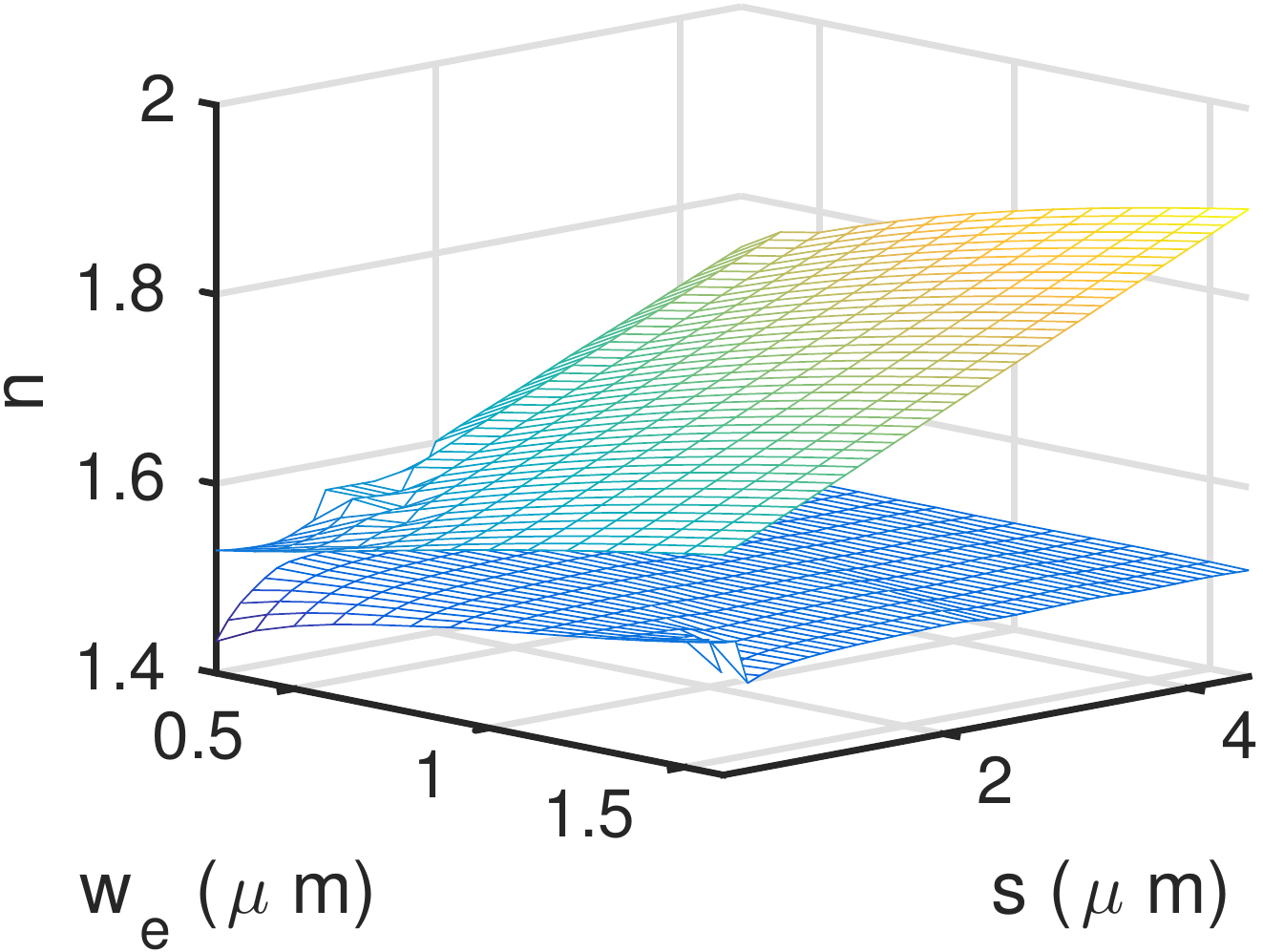}\\
{ (c)}&{(f)}\\
\includegraphics[width=1.7in]{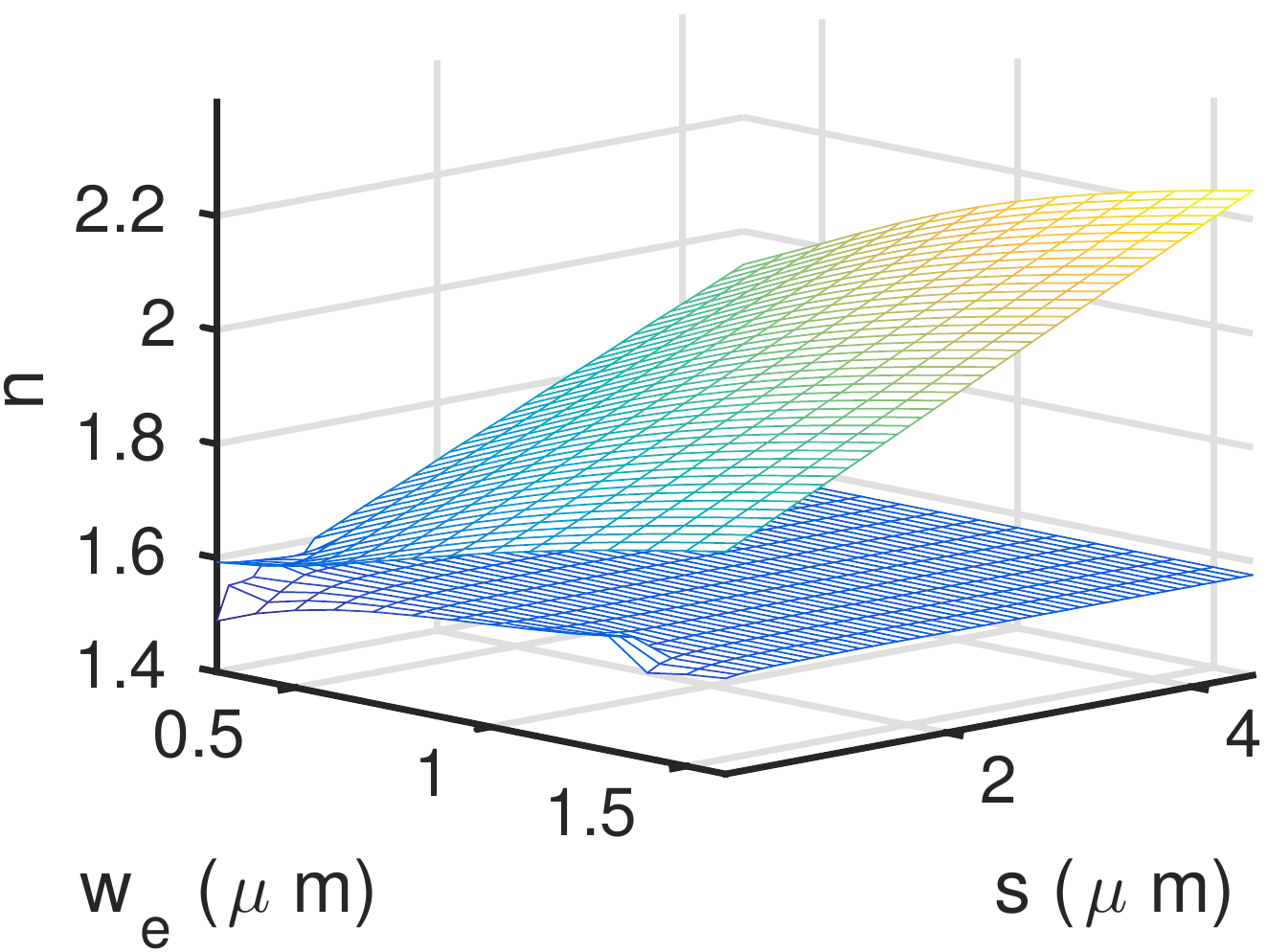}&\hspace*{-0.75cm}
\includegraphics[width=1.7in]{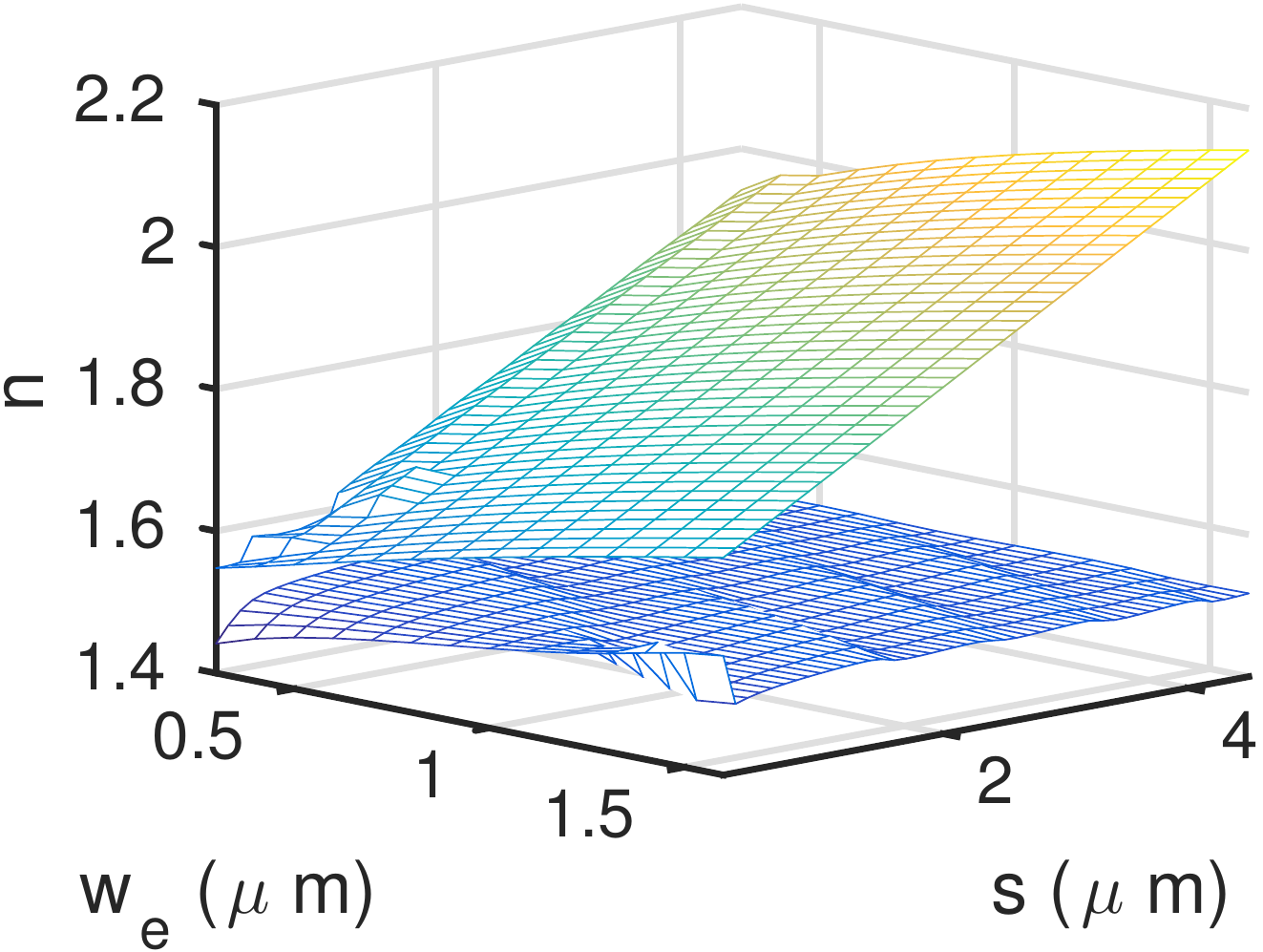}
\end{tabular}
\caption{Real part of the modal effective indices $n_r$ in curved asymmetric couplers for TE ($(a)$ to $(c)$) and TM ((d) to (f)) polarized fields. $R=50$ $\mu$m for $(a)$ and $(d)$, $R=25$ $\mu$m for (b) and (e), and $R=15$ $\mu$m for (c) and (f). }
\label{fig::nr}
\end{figure}

\section{Modal analysis of curved asymmetric coupled waveguides}

The dispersion surfaces of the real modal effective indices of quasi-TE (left column) and quasi-TM (right column) modes are shown in Fig.  \ref{fig::nr}.  Three different values of the bend radii are considered $R=50$ $\mu$m (top), $R=25$ $\mu$m (middle) and $R=15$ $\mu$m (bottom).  

\begin{figure}[H]
\centering
\begin{tabular}{cc}
\begin{tabular}{c}
{ (a)}\\
\includegraphics[width=1.6in]{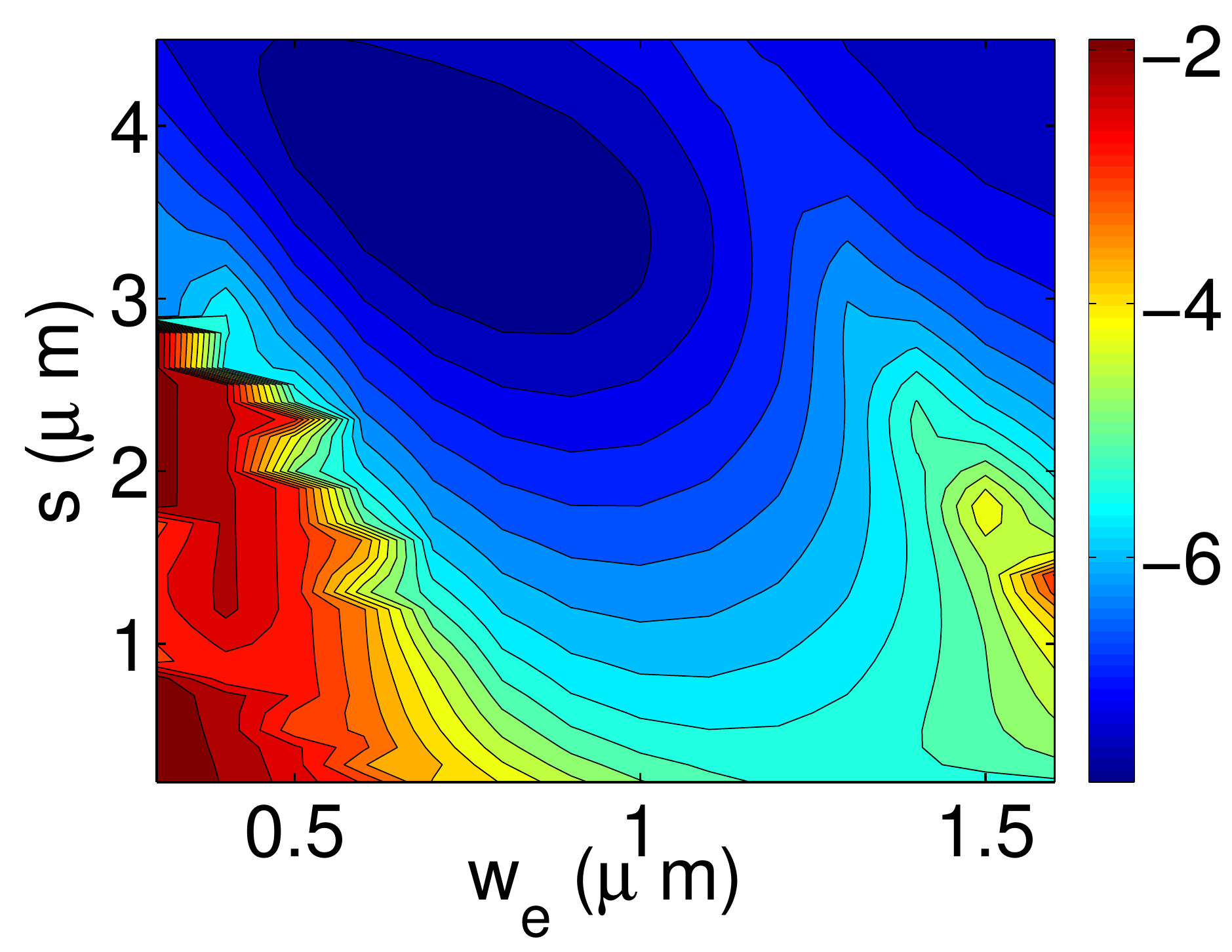}\\
{ (b)}\\
\includegraphics[width=1.6in]{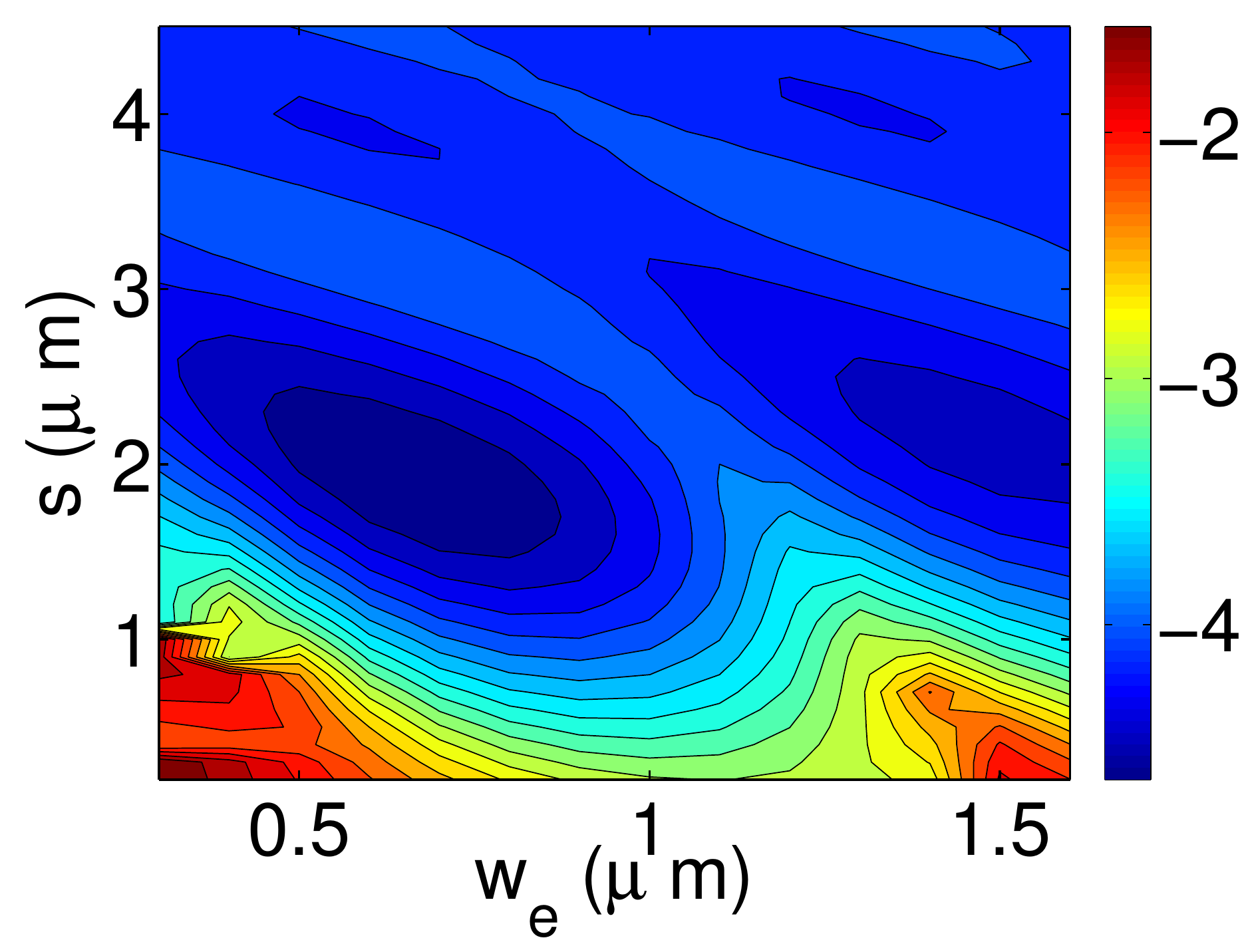}\\
{ (c)}\\
\includegraphics[width=1.6in]{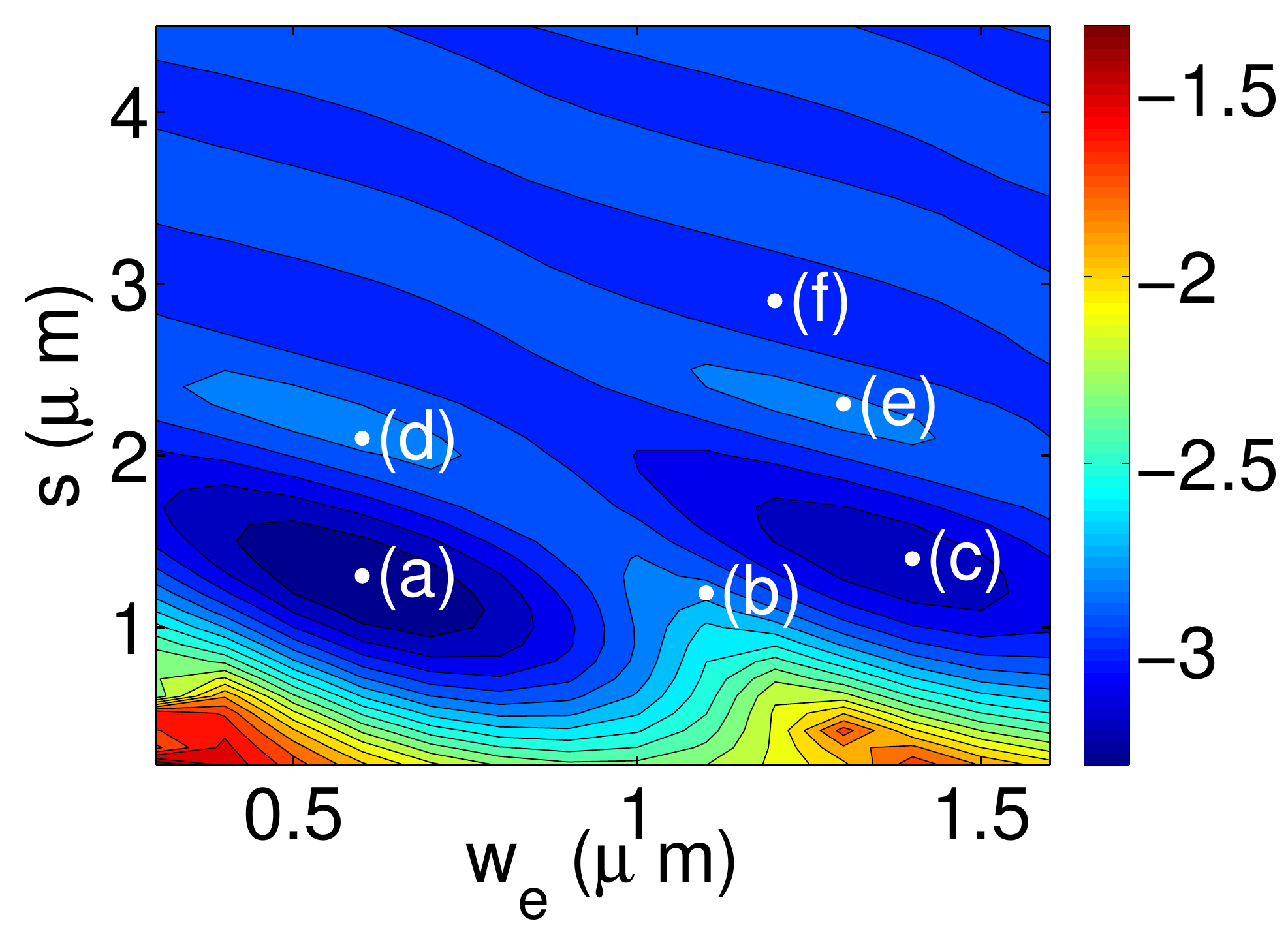}
\end{tabular}&
\hspace*{-0.75cm}
\begin{tabular}{c}
{ (d)}\\
\includegraphics[width=1.6in]{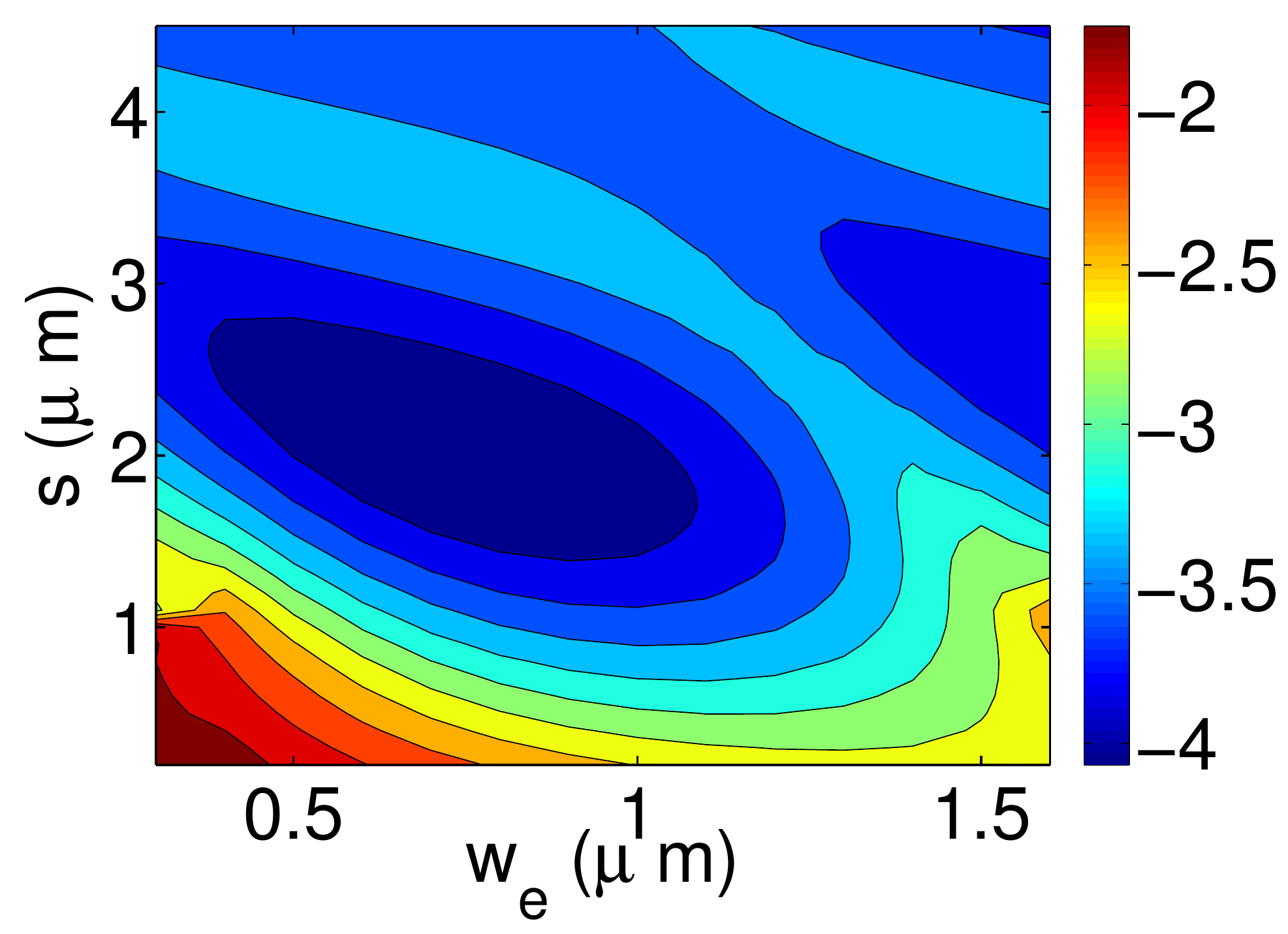}\\
{ (e)}\\
\includegraphics[width=1.6in]{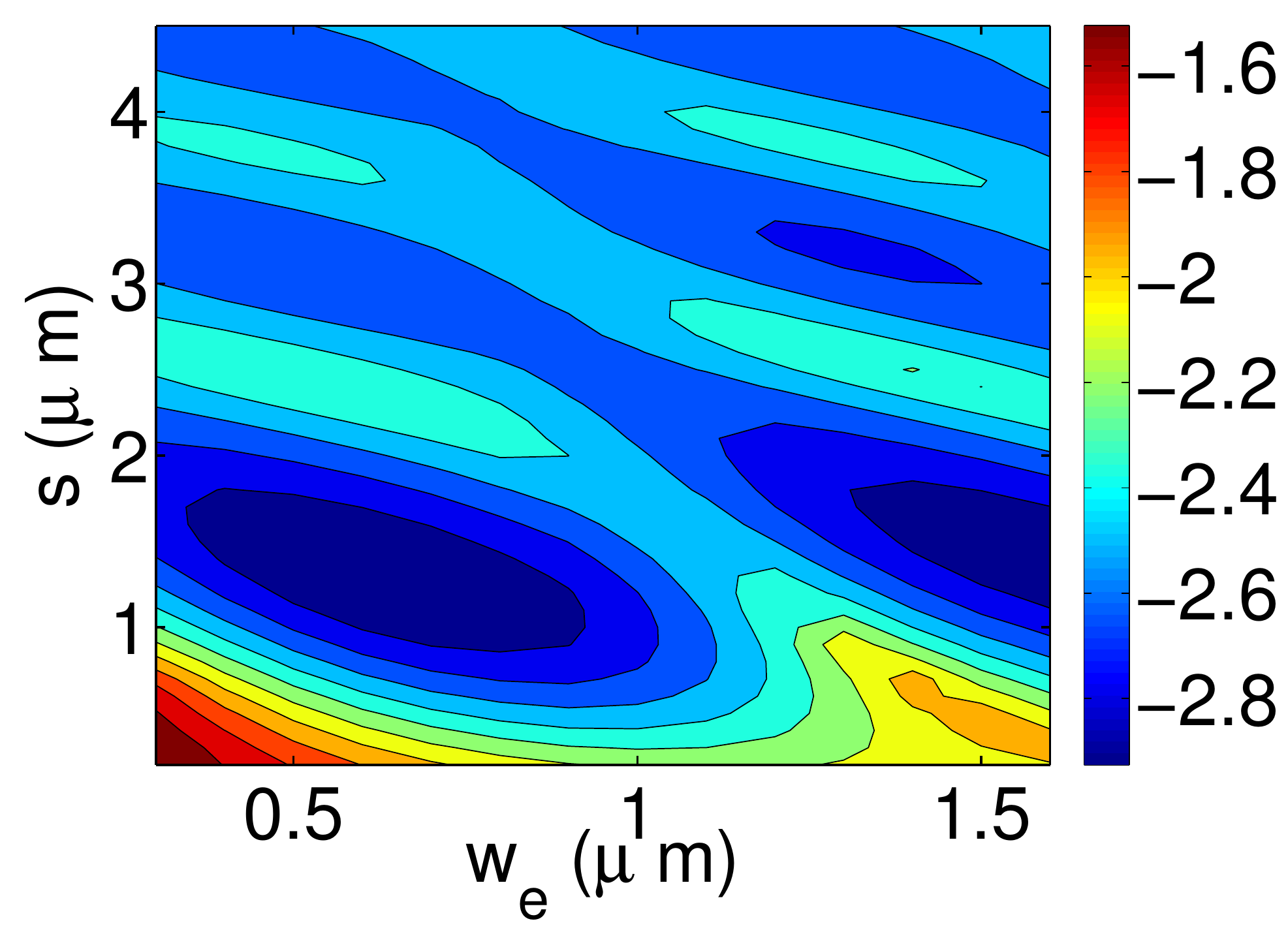}\\
{ (f)}\\
\includegraphics[width=1.6in]{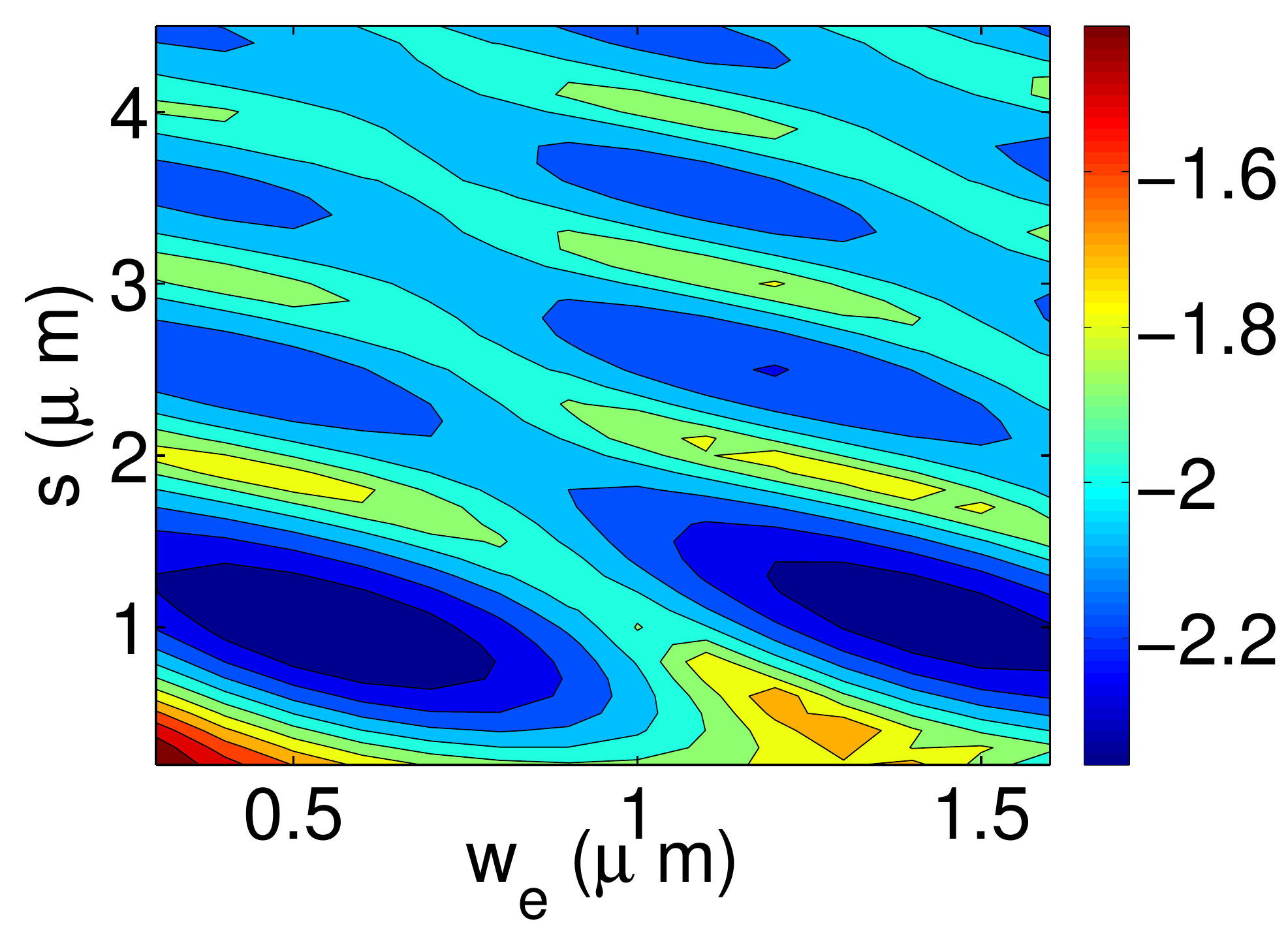}
\end{tabular}
\end{tabular}

\caption{Contours of $log_{10}(n_i)$ of the lower sheet mode for $R=50$ $\mu$m, $R=25$ $\mu$m and $R=15$ $\mu$m corresponding, respectively, to (a), (b) and (c) plots for the TE field and to (d), (e) and  (f) for the TM field. $n_i$ is the imaginary part of the modal index.  Results correspond to full vector 3D calculations \cite{krause}.}
\label{fig::ni}
\end{figure}

If we follow the results in Fig. \ref{fig::nr} from top to bottom, we observe how the curve that defines the degeneracy condition at large $s$, which is $w_e=1$ at $R\to\infty$, is progressively shifted as $R$ decreases.  Accordingly, the range of values of $w_e$ and $s$ for which the lower sheet mode is localized at the inner waveguide grows with the curvature.  In fact, the mode fields of the lower sheet solutions of the dispersion diagrams in Fig. \ref{fig::nr}, both for TE and TM polarizations, are localized at the main waveguide for most values of $w_e$ and $s$.  This is evidenced by the fact that their effective indices surface is very close to the plane defined by the effective index of the isolated main waveguide.  Therefore, for sufficiently large values of $s$, the coupling to the outermost waveguide is negligible and only the lower sheet has to be considered for the calculation of the loss in the curved structure.   

We also observe how an increase of $1/R$ results in larger growths  with $s$ and/or $w_e$ of the effective index of the upper sheet mode (the one that is in almost all the cases localized at the exterior waveguide). It is due to a corresponding increment of the actual bend radius of the exterior waveguide relative to the reference value of $R$ that is taken at the core center of the interior waveguide.  This produces an apparent increase of the modal effective index, and it is also consistent with the effective increase of the refractive index along the radial distance produced by the curving of the guiding structure.

\begin{figure}[H]
\centering
\begin{tabular}{cc}
\includegraphics[width=1.5in]{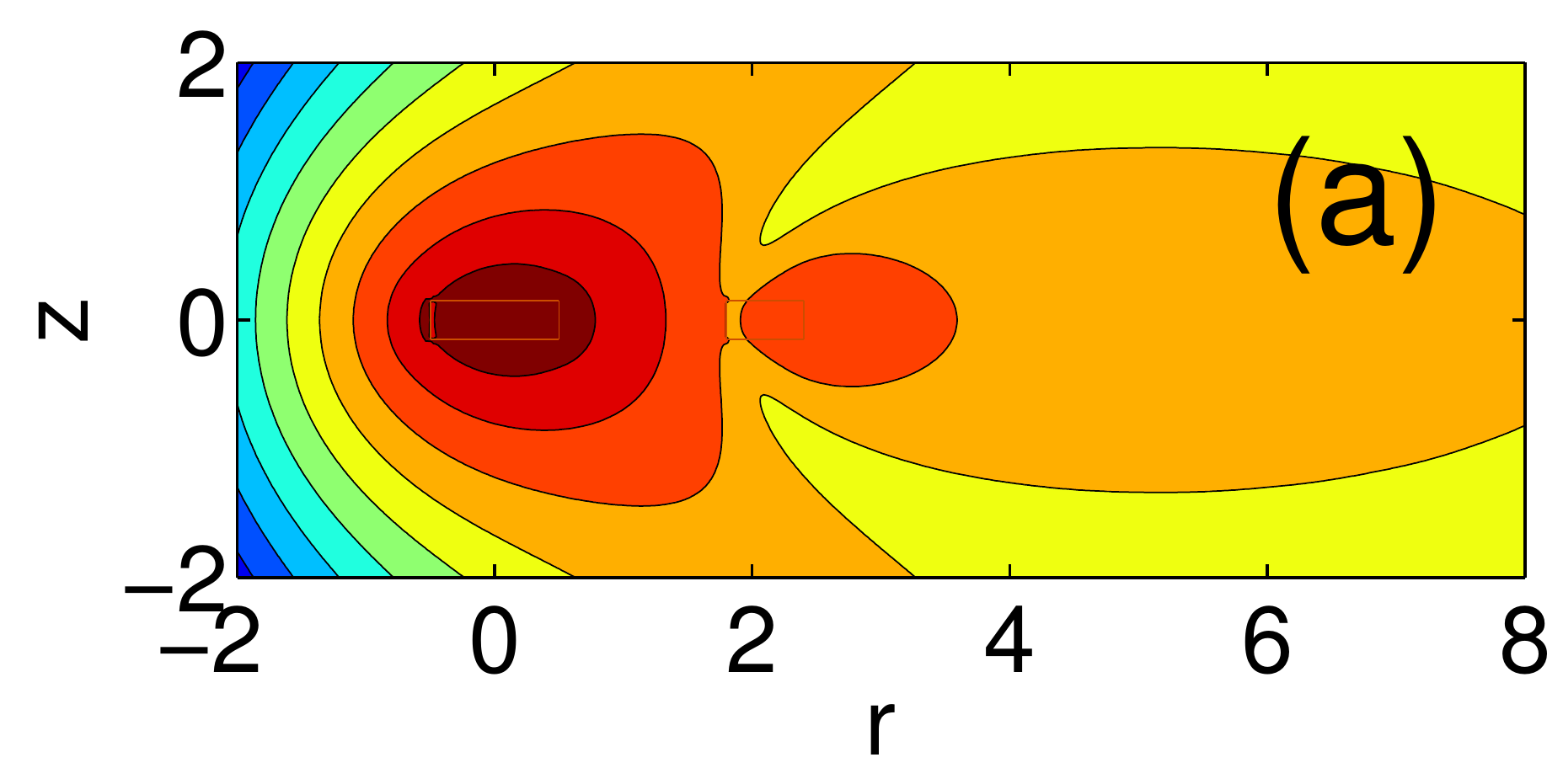}&\includegraphics[width=1.5in]{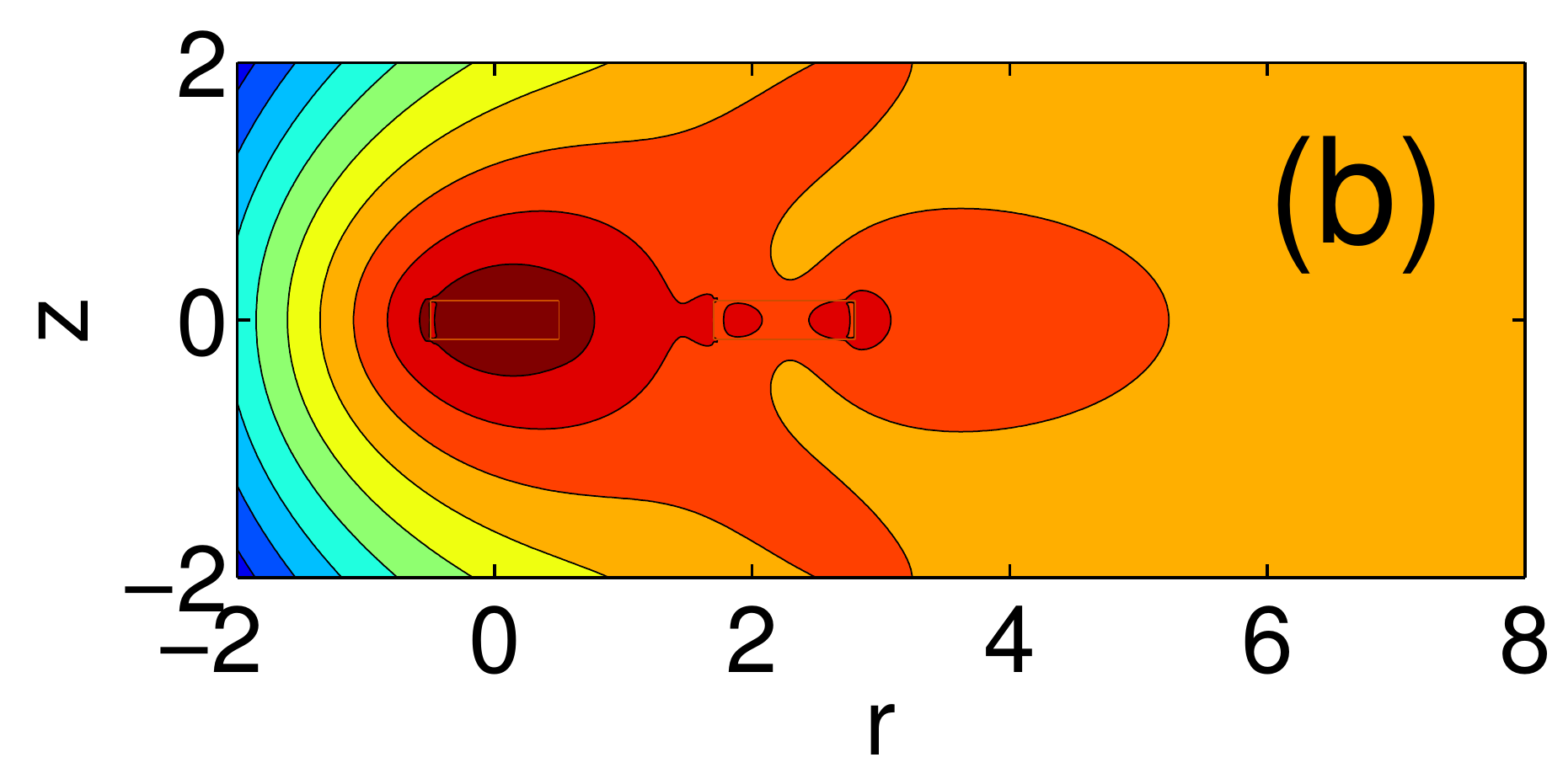}\\
\includegraphics[width=1.5in]{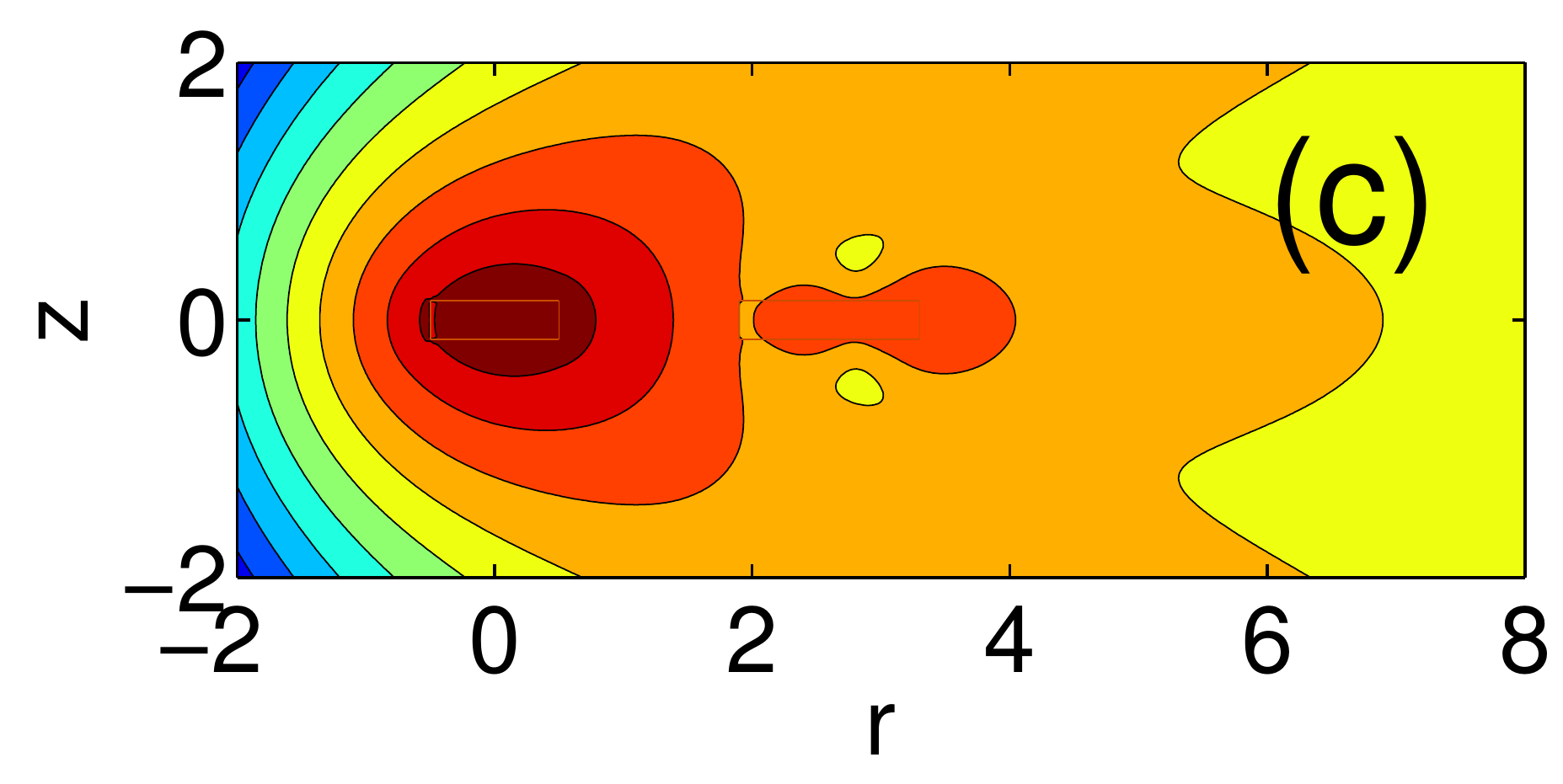}&\includegraphics[width=1.5in]{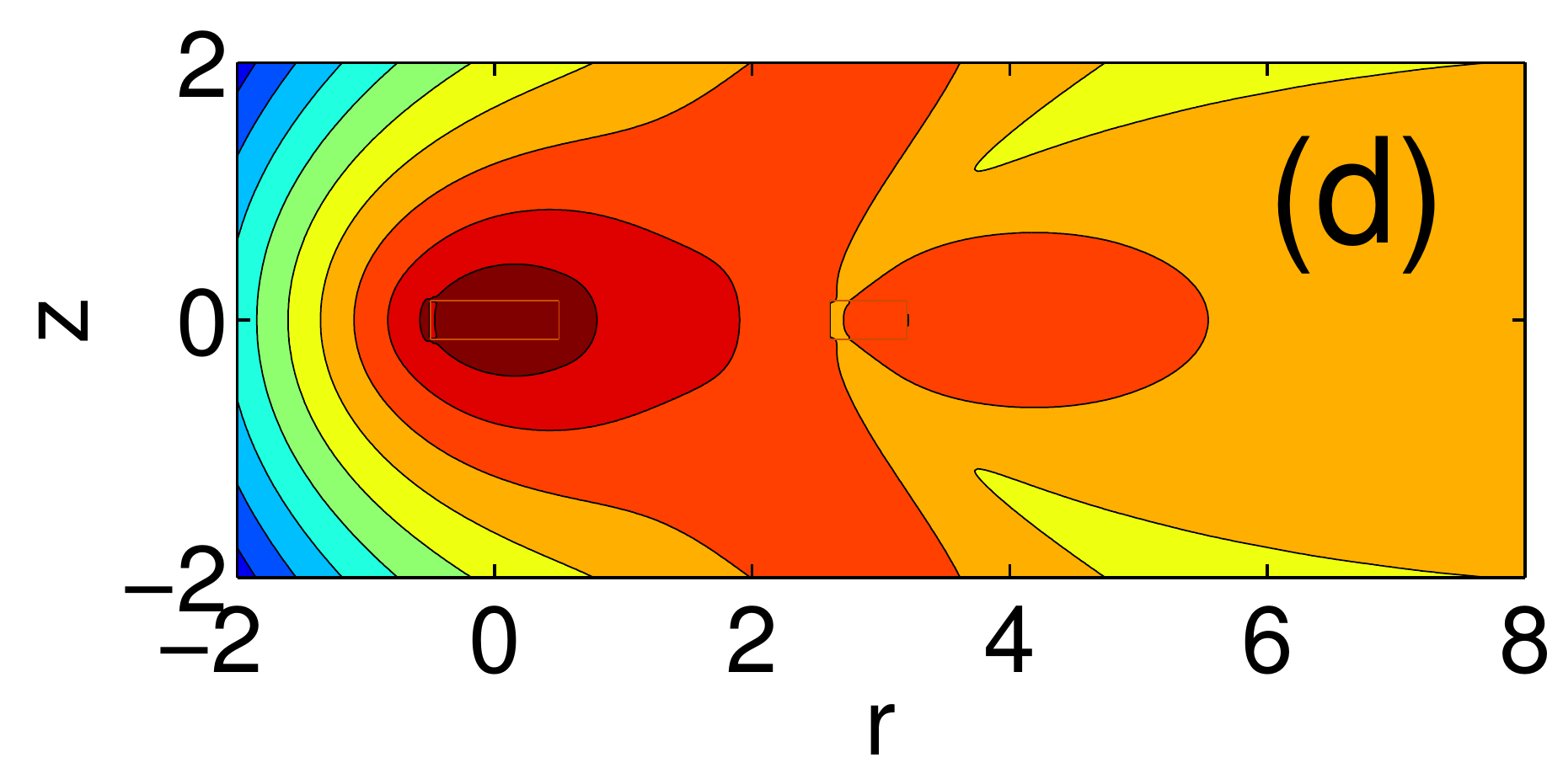}\\
\includegraphics[width=1.5in]{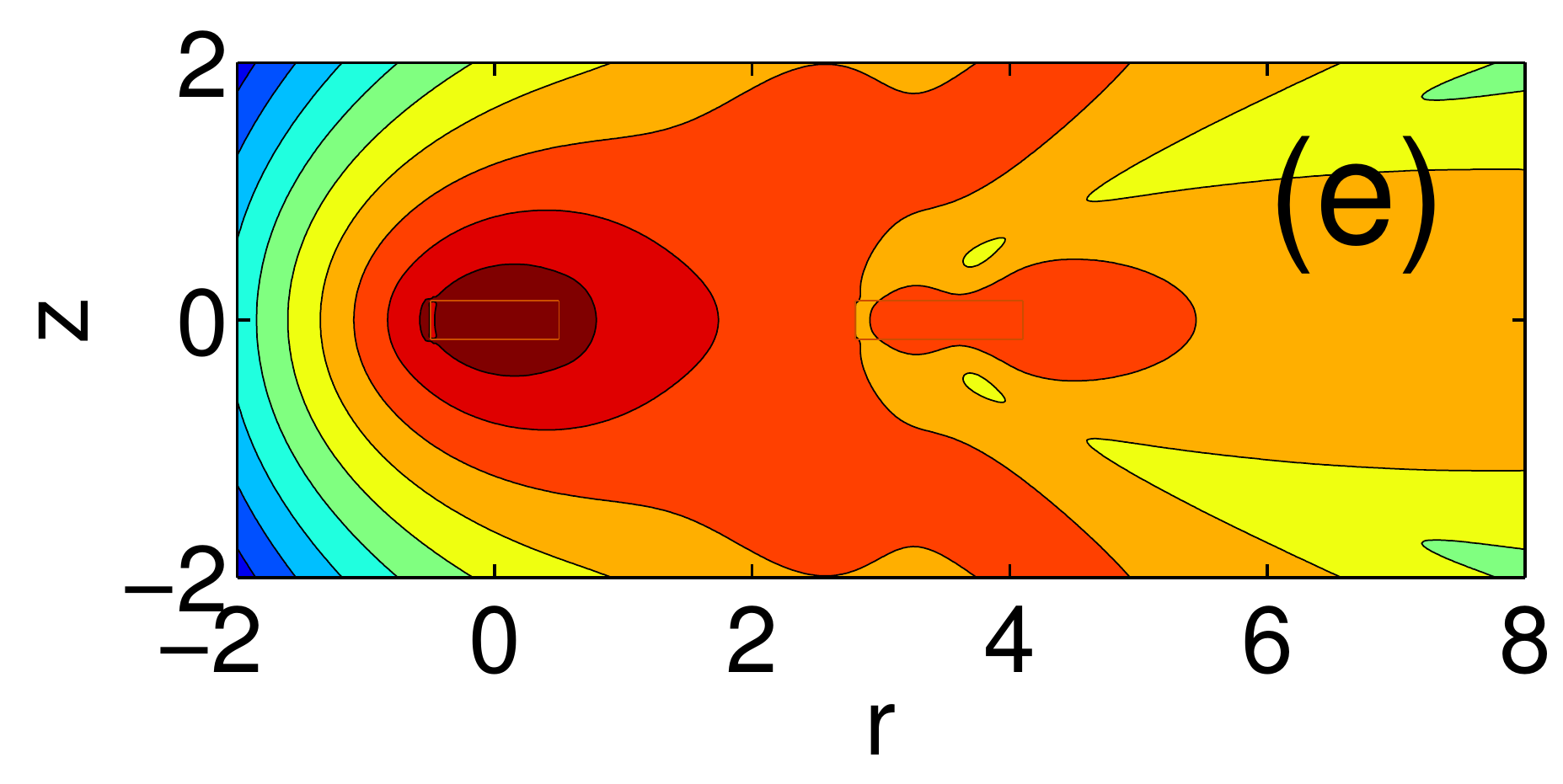}&\includegraphics[width=1.5in]{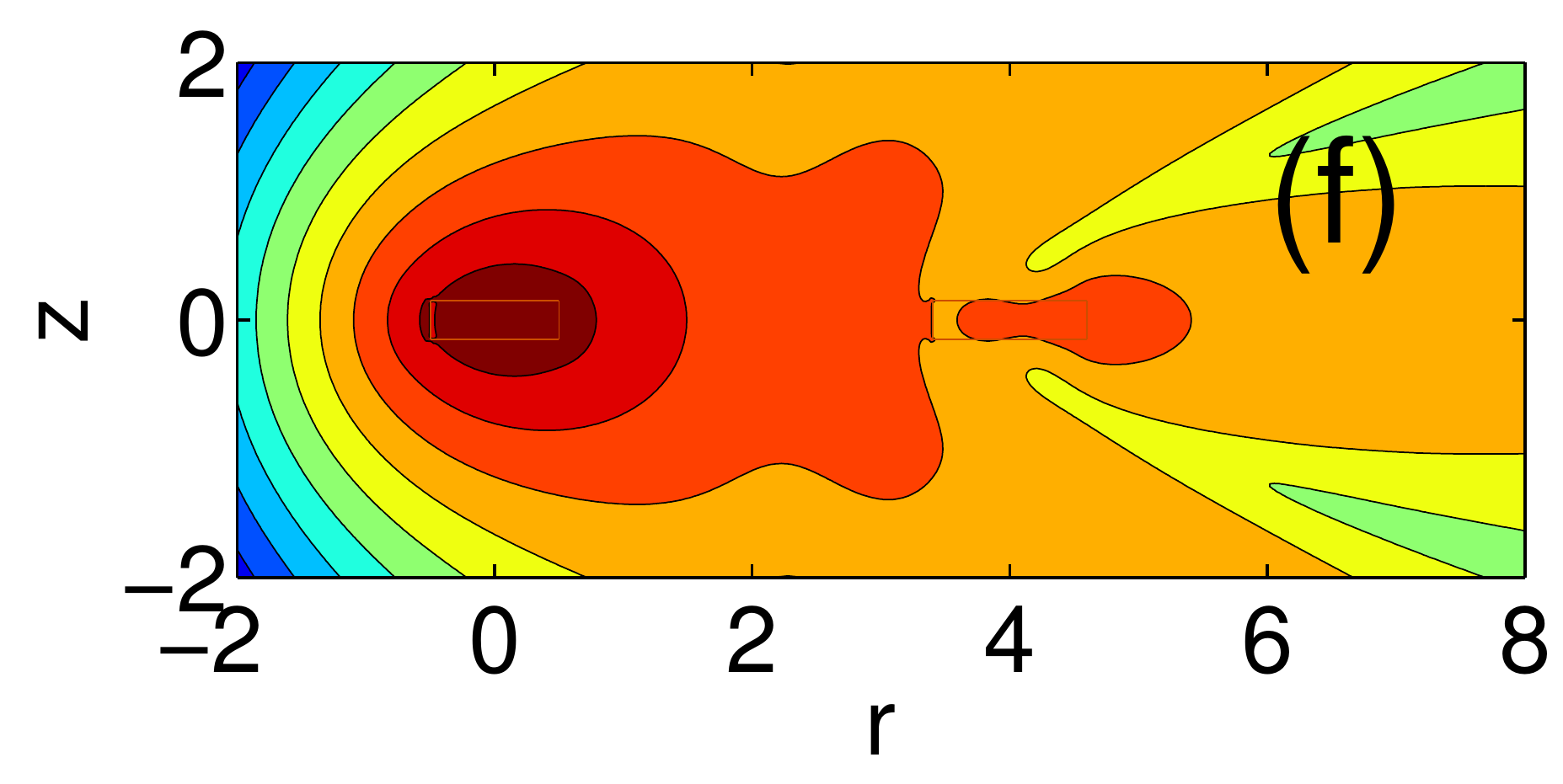}\\
\end{tabular}
\caption{Logarithmic transverse intensity contours for the lower sheet quasi-TE polarized mode in a bent asymmetric structure with $w_e$ and $s$ as displayed in Fig. \ref{fig::ni}(c).}
\label{fig::logintensity}
\end{figure}

The modal losses for the lower sheet, as obtained with full 3D vector calculations, expressed as $\log_{10}(n_i)$ are shown in figure \ref{fig::ni}.  Again, results for quasi-TE (left column) and quasi-TM (right column) polarizations and three values of the bend radius: $R=50$ $\mu$m (top), $R=25$ $\mu$m (middle) and $R=15$ $\mu$m (bottom) are presented.  The results display some expected general features:  In general, radiation loss grows with curvature.  Also, the less-confined quasi-TM modes display larger radiation loss values than the quasi-TE modes for equal bend radii. Besides, a series of resonances and anti-resonances can be observed in the $w_e$ $s$ plane, the latter defining regions where the asymmetric coupler structure can provide a reduction of the radiation loss, as it was described in \cite{chamorro} from 2D FDTD calculations. 

The relative heights (depths) of these resonances (anti-resonances) decrease as either $w_e$ or $s$ increase, therefore defining an optimum radiation quenching region at the first valley in the $w_e$ $s$ plane.  As the bend radius is diminished, the separation between the resonances along the $w_e$ and $s$ coordinates decreases.  This, in turn, produces a shift of the optimal parameters of the exterior ring to smaller values of $w_e$ and $s$ as the curvature increases.

The domains of enhanced and quenched radiation loss are related with radial resonances that determine the mode confinement properties.  This is illustrated in figure \ref{fig::logintensity} where the transverse intensity for the TE polarized lower sheet mode at different values of $w_e$ and $s$ and $R=15$ $\mu$m are shown.  These results correspond to full 3D vector calculations and are displayed in logarithmic scale to provide a more detailed view of the low intensity radiation components.  The specific values of $w_e$ and $s$, associated to either high or low radiation loss points, are marked in Fig. \ref{fig::ni} (c). Cases (a), (c) and (f) correspond to positions close to local minima of radiation loss and (d) and (e) to local maxima.  Case (b) is a transition point. For all the calculated modes, a large degree of localization of the optical field in the main waveguide is found, as expected for the  lower sheet modes.  Also, the field distribution, with a marked local minimum in between the two guides is consistent with the continuation from an odd solution along the radial direction in the  straight coupler.  The relation between the radiation loss resonances and anti-resonances and the modal field distribution is also evident from these results, with the field intensity spreading towards the second waveguide and in two symmetric off-radial directions in the former case and retreating towards the main waveguide core in the latter.

Again, the results in Fig.  \ref{fig::logintensity} witness the intrinsic 3D nature of the radiation in bent waveguides beyond the non-separability evidenced in the curved single core waveguide case of Section \ref{sec::modeling}.  Even thought 2D methods, derived from the EIM, and possibly including further approximations, are very valuable for qualitatively describing the radiation losses, accurate calculation require of a full 3D vector analysis.   Amongst the 2D methods, the solution of the Maxwell in the time domain in the associated 2D domain is free of any further approximation and  provides the most accurate solutions. The comparison of the results presented in Fig. \ref{fig::logintensity} and those obtained with FDTD calculations \cite{chamorro} shows that 2D calculations provide a good qualitative description of the problem and approximate positions for the radiation coupling anti-resonances, even though the deviations from their actual values are significant.  

As shown in Fig. \ref{fig::radiacion}, the imaginary modal indices in a single bent waveguide are $n_i=1.248\times 10^{-3}$ and $n_i=8.545\times 10^{-5}$ for $=15$ $\mu$m and $=25$ $\mu$m, respectively.  With the coupled section, the lowest values of $n_i$ calculated at each $R$ are  $n_i=4.959\times 10^{-4}$ and $n_i=2.330\times 10^{-5}$.  For the propagation along a full circular section of $2\pi$ rad, the respective reduction factors of the radiated power are $0.56$ and $0.92$ at  $=15$ $\mu$m and $=25$ $\mu$m, respectively. 

\section{Spatial transients}

The former study of the modal properties of bent coupled waveguides alone is adequate to evaluate the radiation loss for a sufficiently large propagation distance in a simple structure. Nevertheless, the transmission lengths involved are always limited and there exist different discontinuities that contribute to the spatial transients and can affect the performance of a radiation reduction coupled section under specific operation conditions.  Therefore, the study of the properties of the field evolution along the device is also a requisite in the design and evaluation of radiation quenching structures. 

\begin{figure}[H]
\centering
\includegraphics[width=3in]{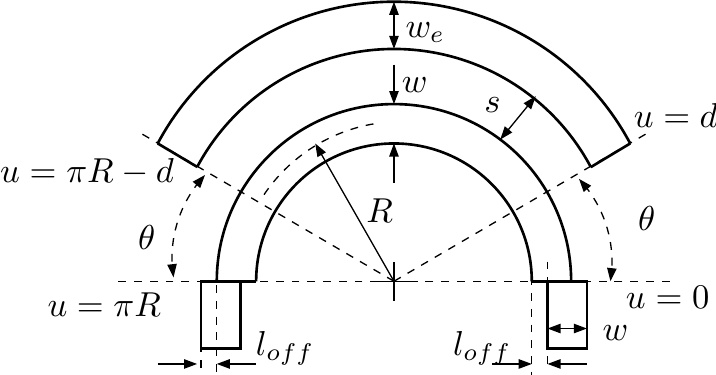}\\
\caption{Geometry used for the study of the spatial transient effects.}
\label{fig::geotr}
\end{figure}

Fig. \ref{fig::geotr} depicts the geometry of the problem addressed in this section, which consists of straight input and output waveguides and a curved section partially arranged in a radiation quenching configuration.  $\theta\ne 0$ is assumed as a result of geometry restraints arising, for instance, from the requirement of coupling to an external circuit.  The evolution of the optical field is calculated in the range from $u=0$ to $u=\pi R$, where the curved waveguide is coupled again to the straight output waveguide. We  restrict our analysis to TE-polarized field, which display a smaller level of radiation, and to a radius of $R=15$ $\mu$m.

The characteristics of the evolution of the optical field following a transition between a straight and a bent waveguide section  depend on the launching conditions.    Notably, the coupling loss can be reduced by simply introducing a lateral shift of the axis of the two waveguides \cite{kitoh} in order to improve the overlap between the modal field in the straight input waveguide and in the curved section, where the distortion of the field profile includes a displacement of the maximum intensity in the radial direction (see Fig. \ref{fig::radiacion}).  This also has a relevant effect on the transient field evolution.  

\begin{figure}[H]
\centering
\includegraphics[width=2.8in]{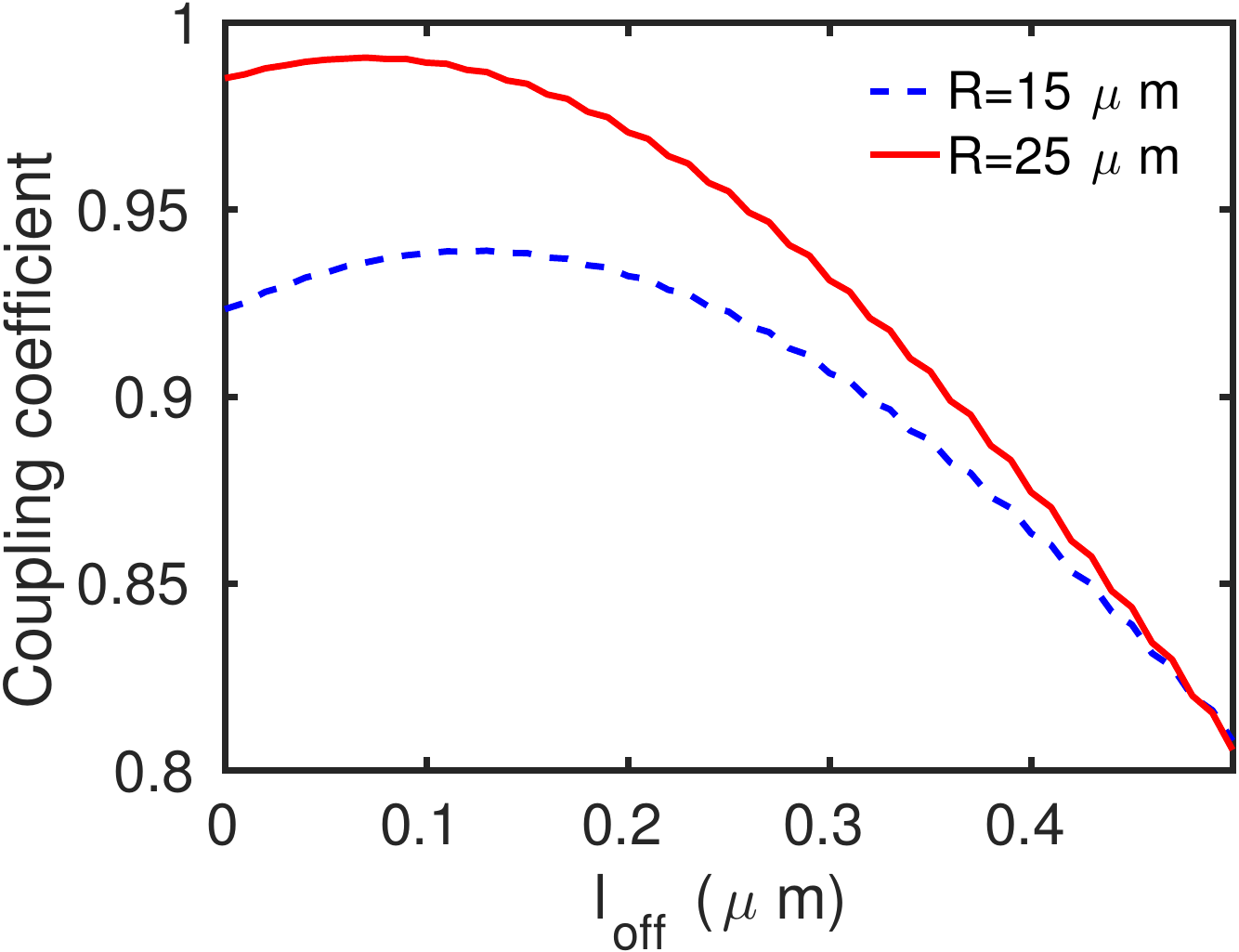}\\
\caption{Coupling coefficient between the modal fields of a straight and bent waveguide as a function of the lateral offset $s$. Data for both $15$ $\mu$m and $25$ $\mu$m radii are displayed. Dotted lines show the results in the absence of the exterior ring.}
\label{fig::offset}
\end{figure}

\begin{figure}[H]
\centering
\begin{tabular}{cc}
\hspace*{-0.25cm}\includegraphics[width=1.8in]{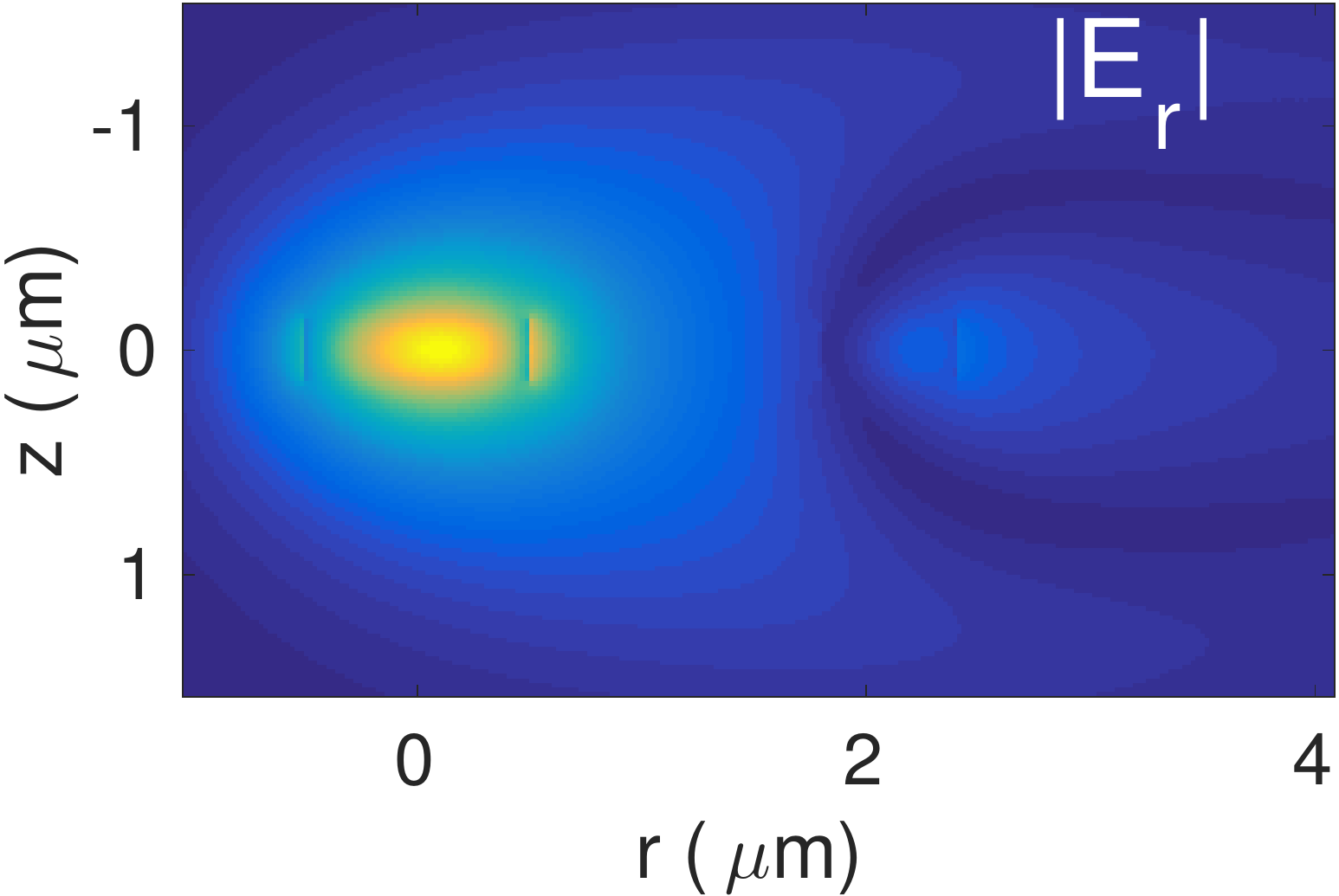}&\hspace*{-0.75cm}\includegraphics[width=1.8in]{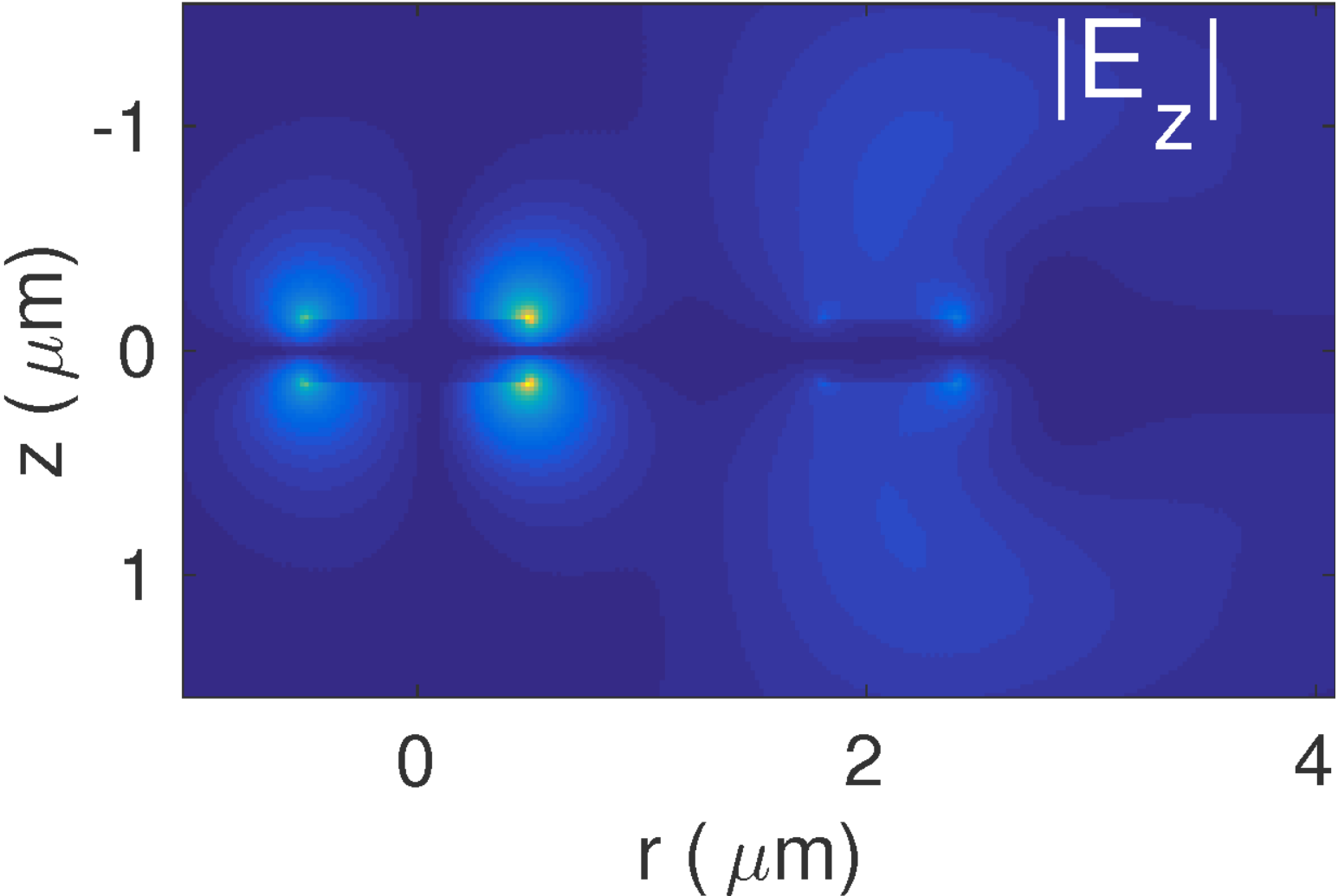}
\end{tabular}
\caption{$r$ and $z$ polarized electric field intensity (left and right plots, respectively) at $u=30$$\mu$m. }
\label{fig::BPMcampo}
\end{figure}

\begin{figure}[H]
\centering
\begin{tabular}{cc}
\hspace*{-0.75cm}\includegraphics[width=1.7in]{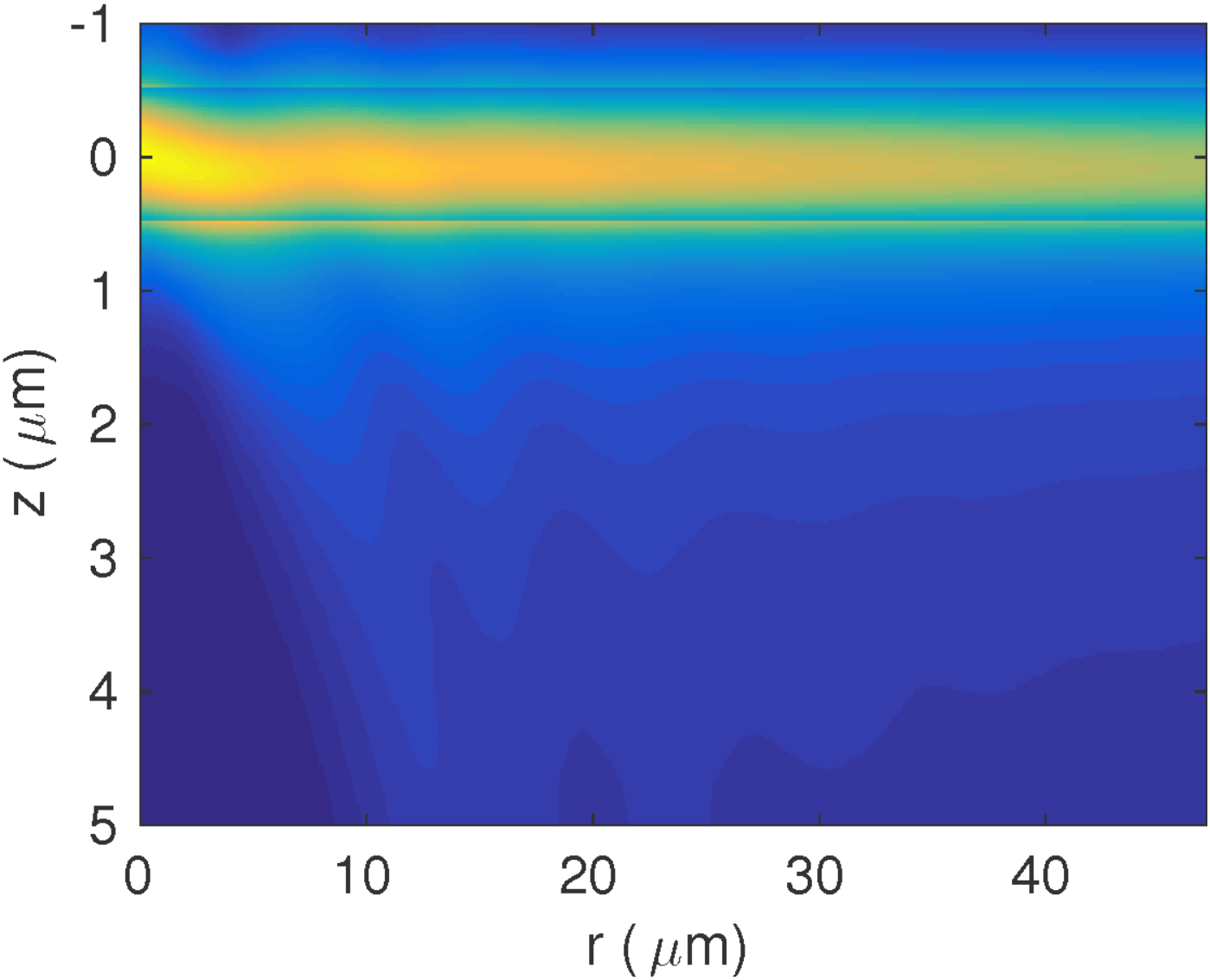}&\hspace*{-0.75cm}\includegraphics[width=1.7in]{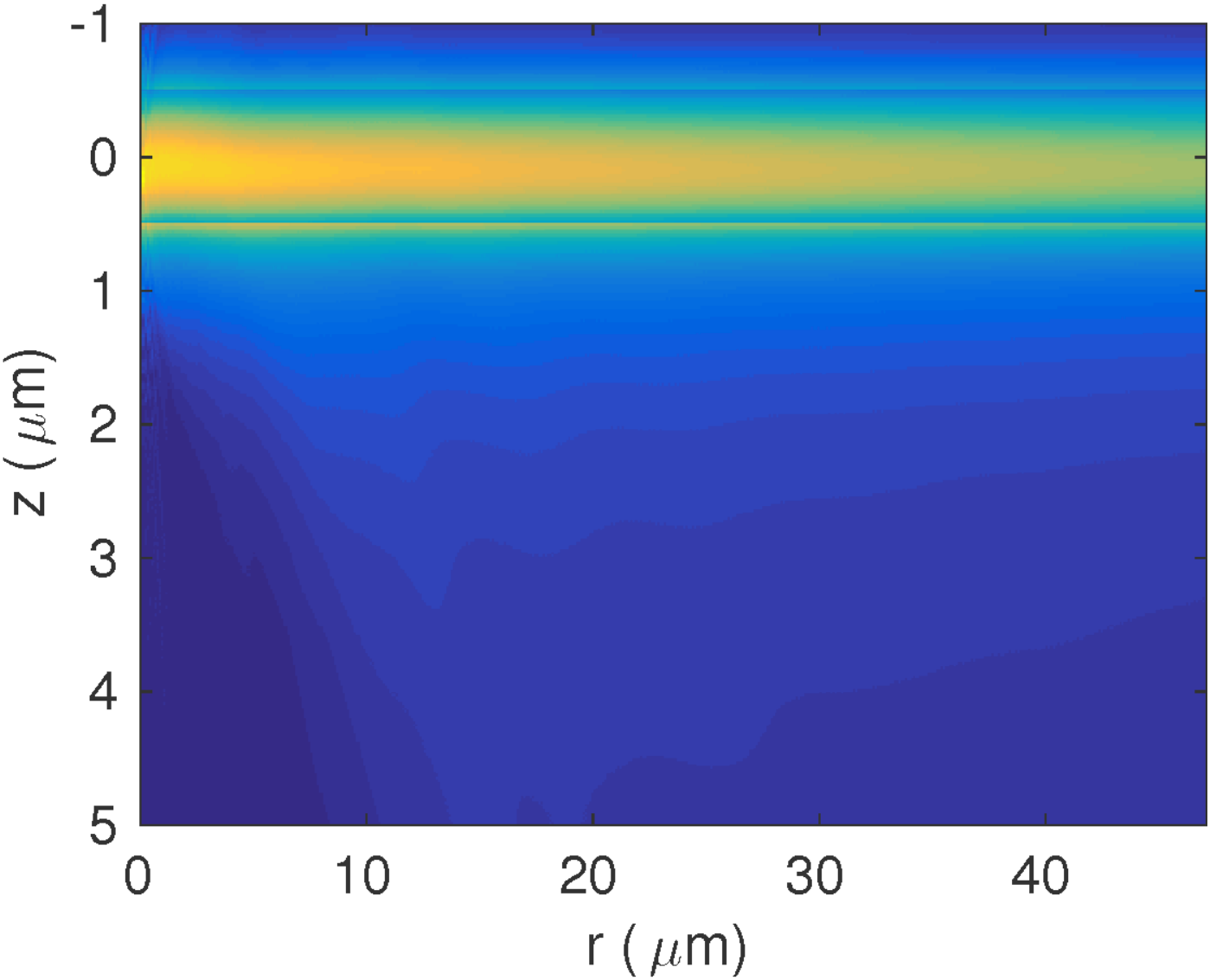}\\
\hspace*{-0.75cm}\includegraphics[width=1.7in]{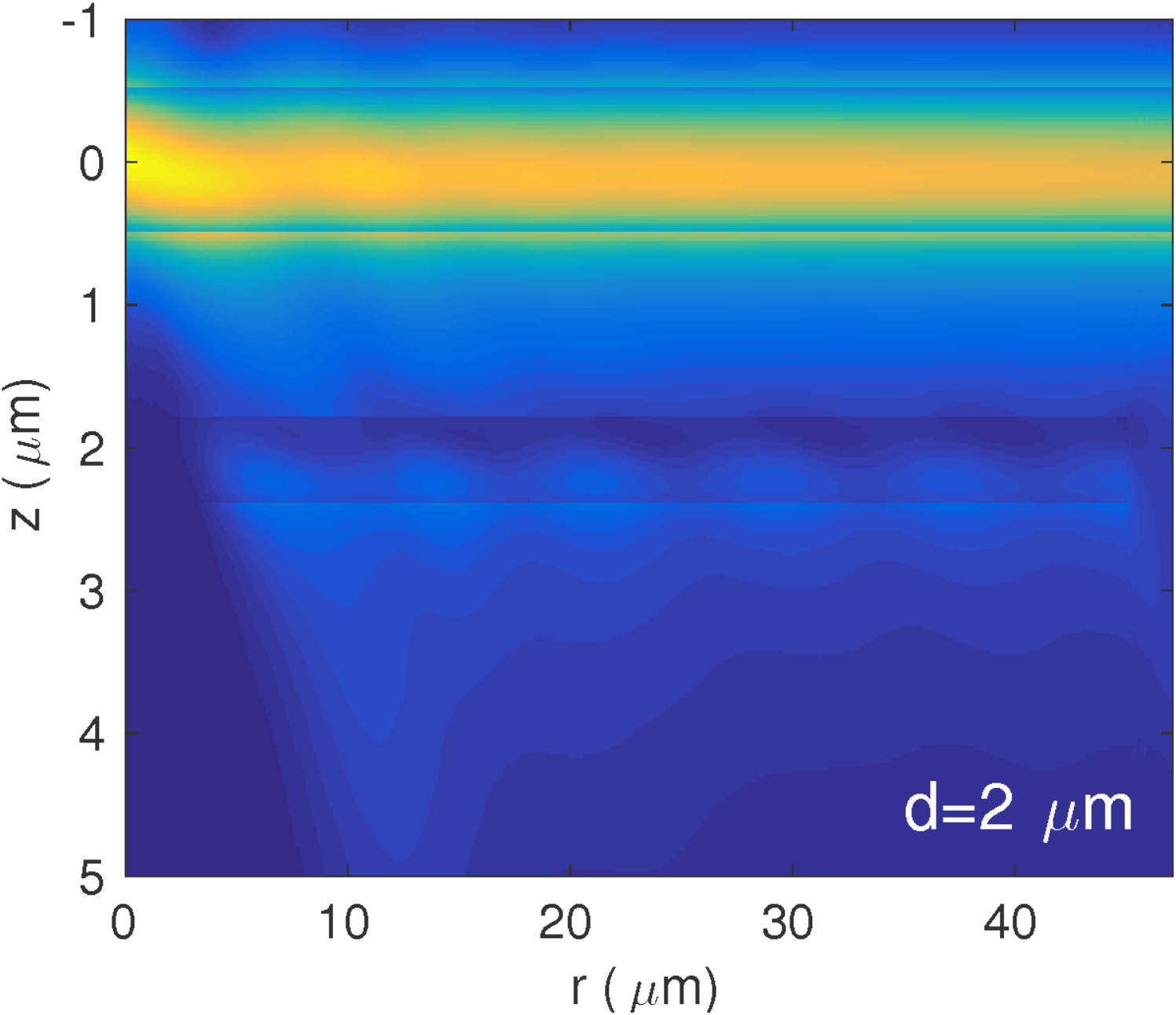}&\hspace*{-0.75cm}\includegraphics[width=1.7in]{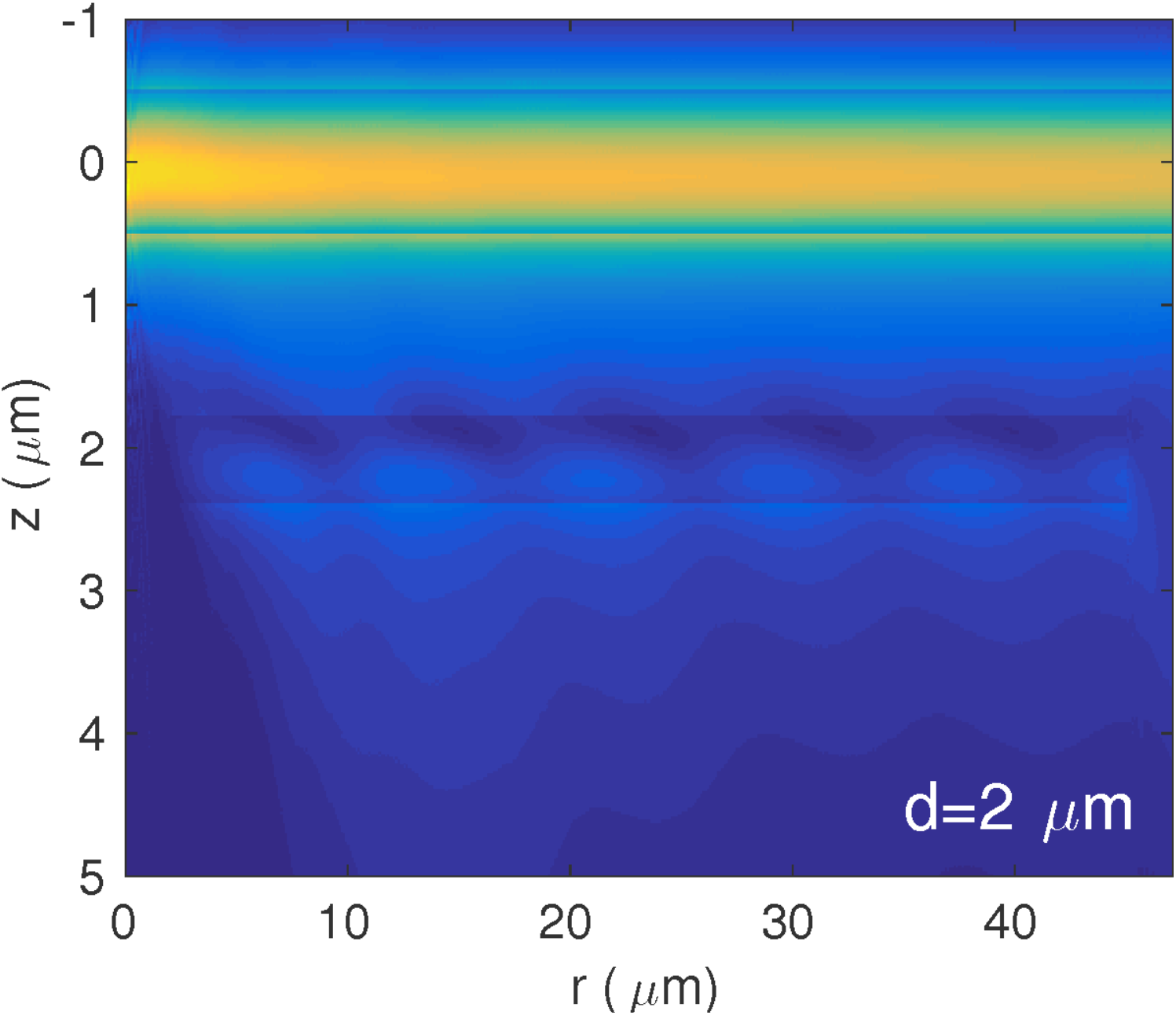}\\
\hspace*{-0.75cm}\includegraphics[width=1.7in]{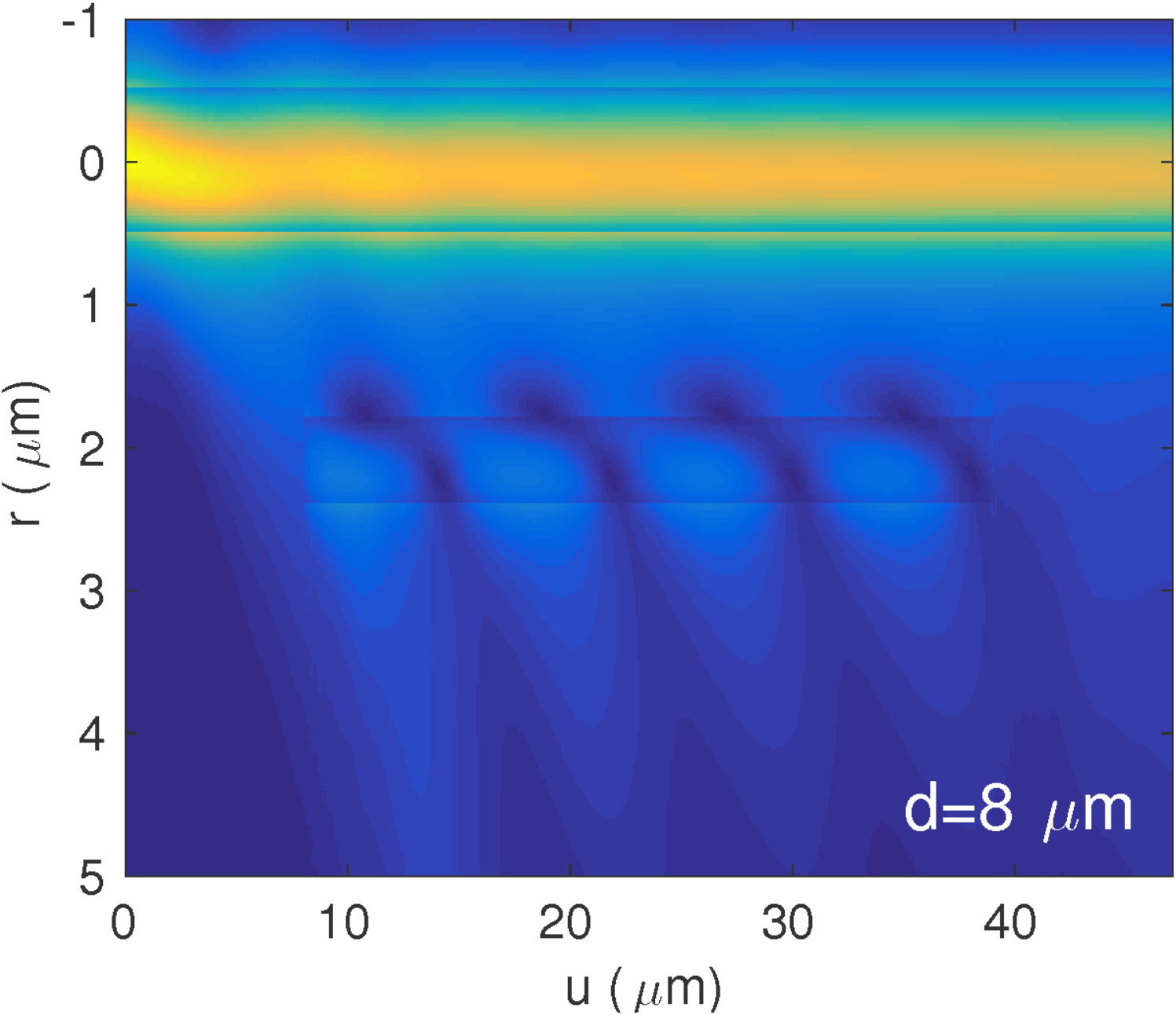}&\hspace*{-0.75cm}\includegraphics[width=1.7in]{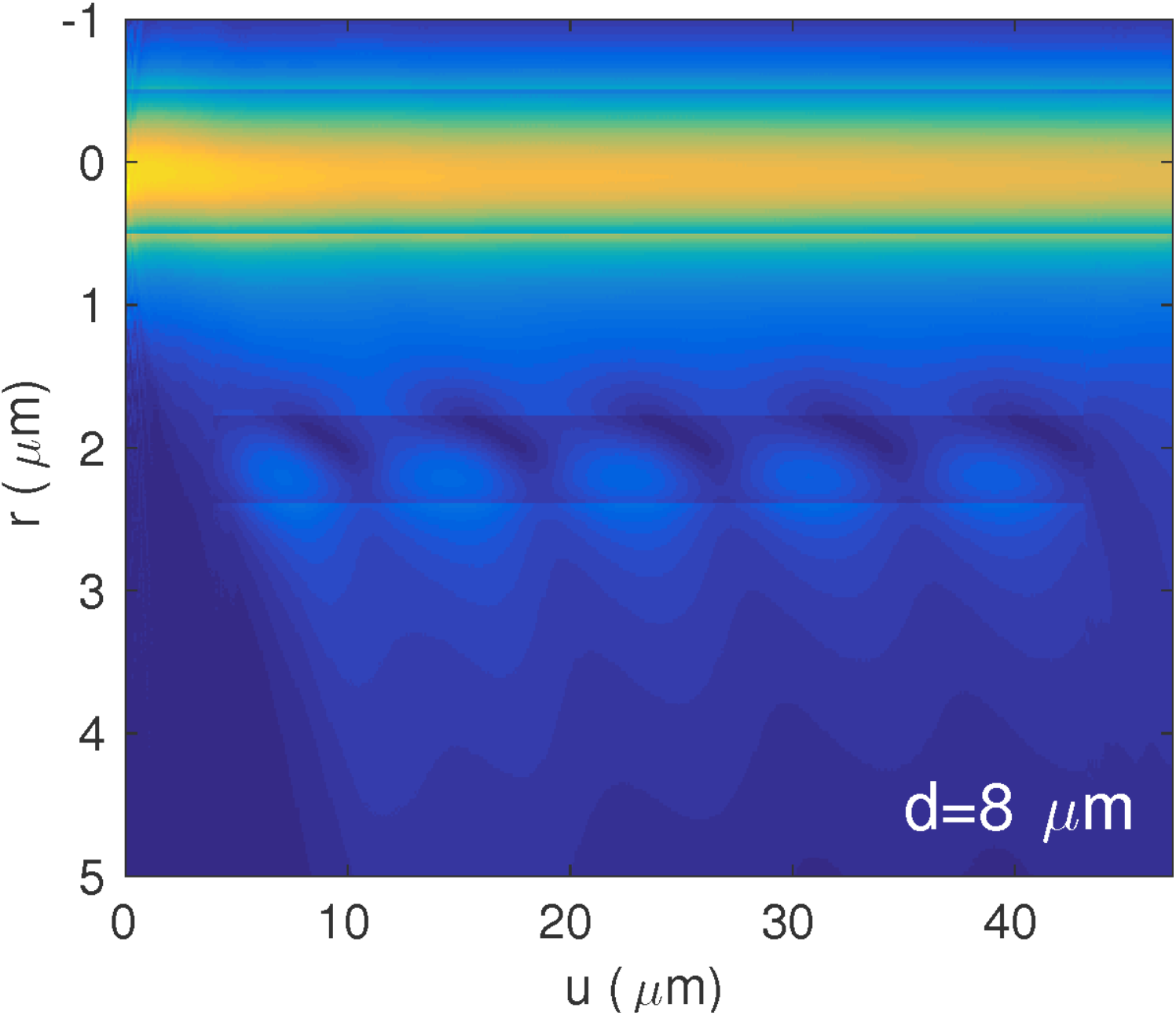}\\
\hspace*{-0.75cm}\includegraphics[width=1.7in]{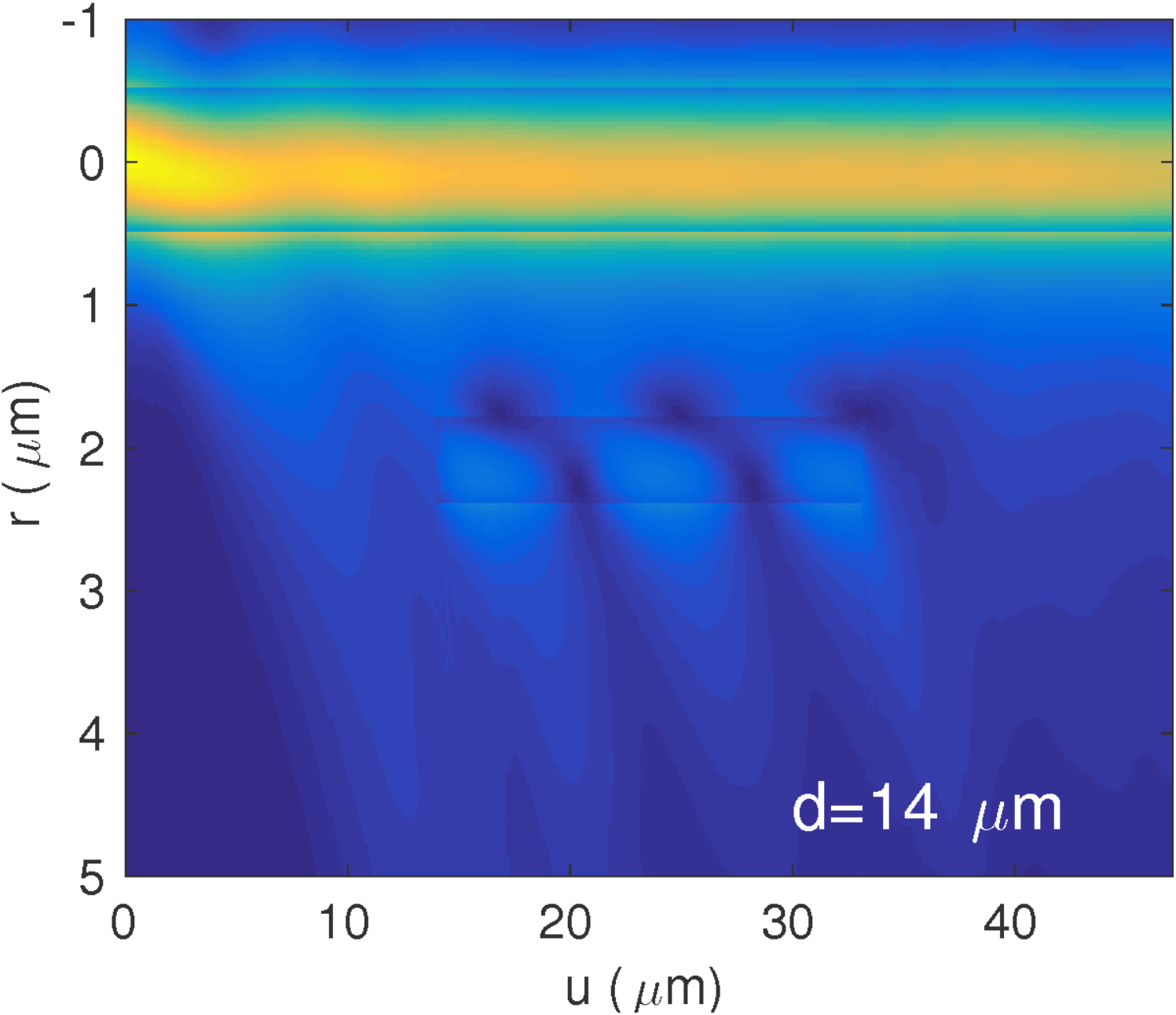}&\hspace*{-0.75cm}\includegraphics[width=1.7in]{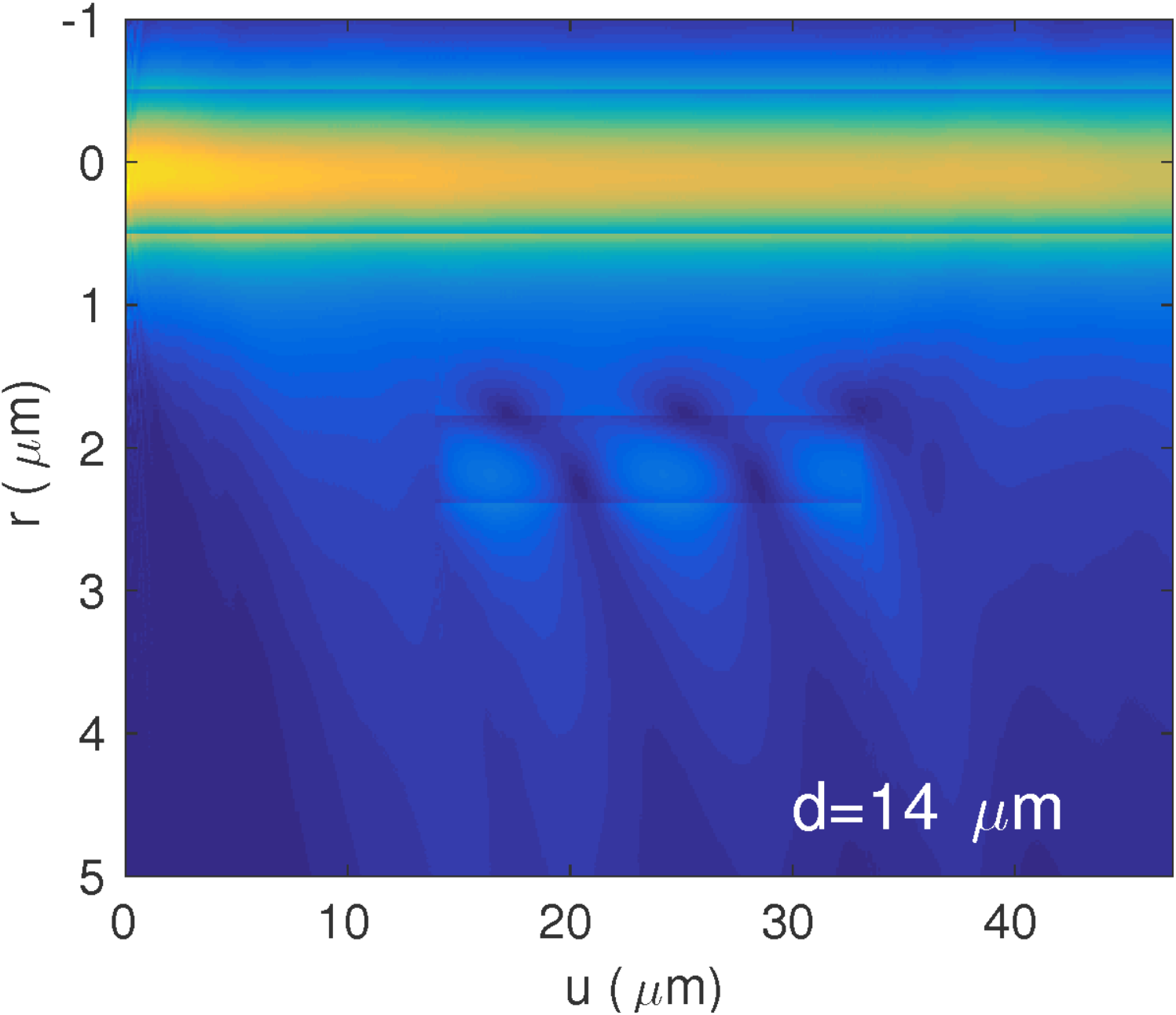}
\end{tabular}
\caption{$r$-polarized electric field intensity in the $r$$u$ plane for a curved waveguide without the exterior ring (top) and for three different values of $d$. Left column corresponds to $l_{off}=0$ whereas the right column to the optimal displacement from Fig. \ref{fig::offset}.}
\label{fig::BPMrz}
\end{figure}

\begin{figure}[H]
\centering
\includegraphics[width=3in]{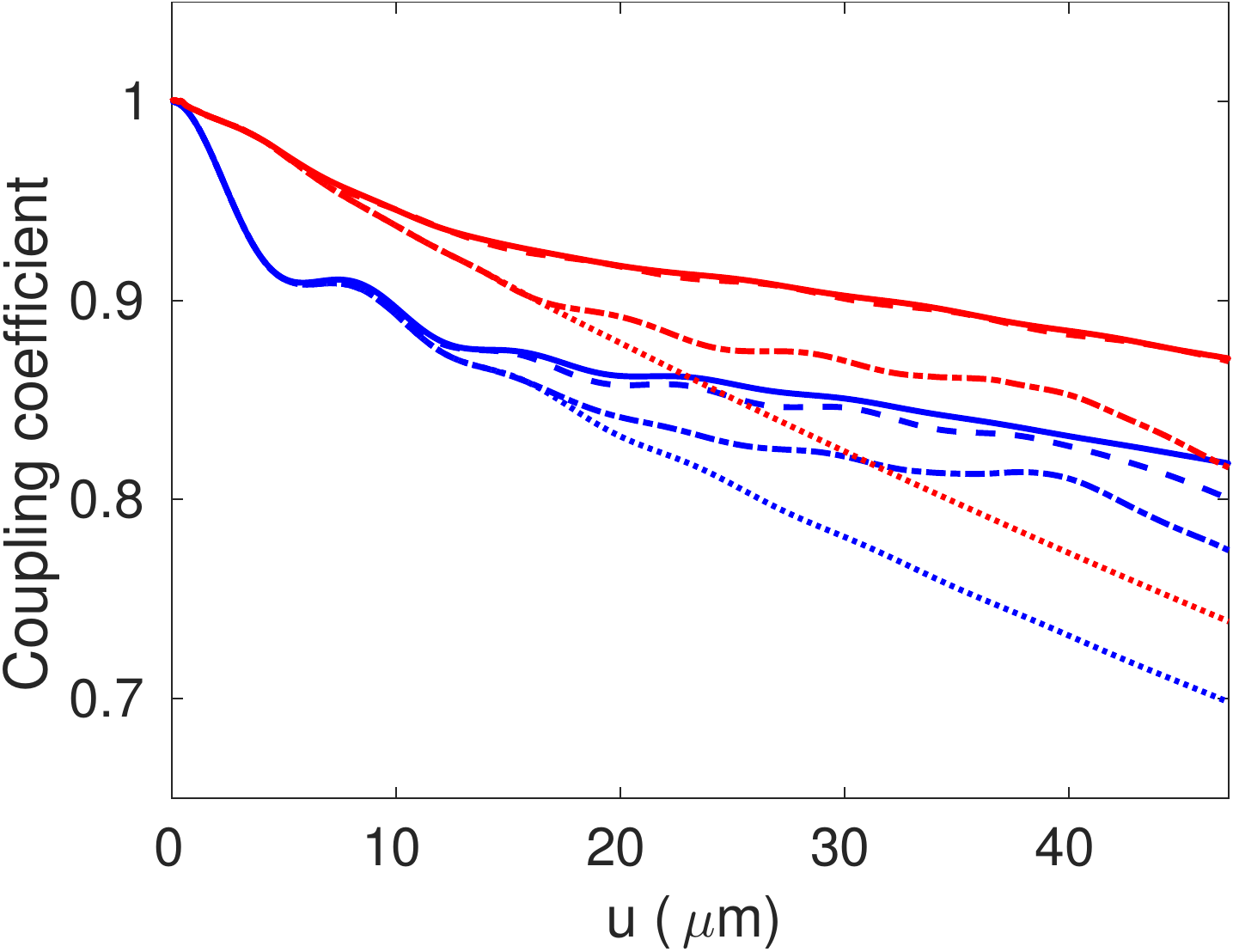}
\caption{Projection of the propagating field onto the modal profile of the input/output waveguide.  Blue lines correspond to $l_{off}=0$ and red lines to the optimal value of $l_{off}$. Solid: $d=2$ $\mu$m, dashed:  $d=8$ $\mu$m and dashed-dotted: $d=14$ $\mu$m. Dotted lines display results in the absence of the exterior arc.}
\label{fig::BPMc}
\end{figure}

The coupling coefficients between straight and bent sections in our waveguide geometry are displayed in Fig. \ref{fig::offset}.  They show a significant reduction of the coupling loss at the optimum lateral offset, particularly for $R=15$ $\mu$m.   It is noteworthy that when similar calculations are performed in the 2D case \cite{chamorro}, we find a spiky variation of the coupling coefficient with $l_{off}$.  This is due to the existence of field discontinuities in the lateral direction.  Nevertheless, this effect is barely noticeable in the full 3D calculations due to the non-separability of the optical field.

The results of the full 3D vector beam propagation calculations for the structure of Fig. \ref{fig::geotr} are shown in Figs. \ref{fig::BPMcampo}, \ref{fig::BPMrz} and \ref{fig::BPMc}. We use the minimal attenuation data set from Fig. \ref{fig::ni}: $w_e=600$ nm and $s=1.6$ $\mu$m.  The mode field in the straight waveguide computed using \emph{wgms3d} has been used as initial condition.  The amplitudes of the $r$ and $z$ components electric field intensity at $u=30$ $\mu$m are shown in Fig. \ref{fig::BPMcampo}. Fig. \ref{fig::BPMrz} shows the magnitude of the $r$ component of the electric field intensity in the $rz$ plane.  The results of the top row, without the coupled waveguide, display a much larger attenuation.  The comparison of the results of the left column, which correspond to $l_{off}=0$, with those of the right column that are computed with an optimal $l_{off}=130$ nm (from Fig. \ref{fig::offset}) provides a clear picture of the impact of the lateral offset in order to improve the coupling and to reduce the spatial transients.

A detailed analysis can be derived from the evolution of the projection of the propagating field onto the mode of the input/output waveguide, as plotted in Fig. \ref{fig::BPMc}.  It is noteworthy that a small length reduction of a full extent exterior arc (a small value of $d$) has little effect on the efficiency of the radiation quenching structure.  Qualitatively, this behavior had been previously observed in the 2D calculations \cite{chamorro}.  The results of Fig. \ref{fig::BPMrz} show that there is a shadow region between the launching point and the position where the illumination reaches the outermost waveguide.  So, the edge of the exterior arc can be safely moved within this dark region without significantly altering the field evolution.  The ``safe'' range of values of $d$ increases when a lateral offset is introduced because the reduction of the mismatch at the input results in a broadening of the shadow region. 

In all cases, the implementation of the optimal lateral offset reduces the amplitude of the transient oscillations displayed in Figs. \ref{fig::BPMrz} and \ref{fig::BPMc}, as it is expected from an improvement of the match of the input to the modal field in the curved structure.   Particularly sustained oscillations are found in the coupled waveguide region at larger values of $d$ because the propagating field then feels more abruptly the discontinuity set by the exterior ring.

It is important to note that the accuracy of the beam propagation calculations is limited by the SVEA and the approximations related with the assumption of a large value of $R$.  Mode-solving calculations, on the other hand, are free from these approximations and they also benefit from improved-accuracy finite difference schemes.  Furthermore, these improved schemes permit an accurate placement of the waveguide boundaries, whereas they are set at mid-distance from the two nearest discretization points in the BPM implementation.  In spite of these limitations, beam propagation results still yield reasonable estimates of the propagation losses.  Moreover, the information provided by BPM calculations is most valuable for the design of practical implementations.

\section{Conclusion}

A numerical study of the properties of curved asymmetric coupled waveguides has been presented.  This type of structure had been put forward \cite{chamorro} for the enhancement of the Q-factor of integrated micro-resonators.
The full 3D vector calculations agree qualitatively with the previous 2D studies based on the EIM, and they offer the degree of accuracy required for the design of practical integrated optics implementations.  Two different analyses have been performed: Highly accurate mode calculations provide optimal parameter sets and they are supplemented with the survey of the spatial transients using beam propagation techniques, which permit to complete the designs by determining the remaining parameters like the angular extent of the exterior arc.

Even though the focus of this work has been placed on the reduction of the radiation loss at the anti-resonance regions, the range of potential application of this type of curved asymmetric coupled waveguides is rather wide. Enhanced radiation at radial resonances can be exploited to realize ultra-compact integrated attenuators.  The addition of a physical mechanism to control the refractive index in the exterior arc can be used to implement variable attenuators and/or integrated modulators.  Also, the high differential attenuation for TE and TM polarizations can be used in integrated polarization handling devices \cite{chamorro18}. 

\section*{Acknowledgement}

This work has been was supported by Junta de Castilla y Le\'on, Project No. VA089U16, and the Spanish Ministerio de Economía y Competitividad, Project No. TEC2015-69665-R.

\end{document}